\documentclass[sigconf]{acmart}

\usepackage{array}
\usepackage{acmart-taps} 

\usepackage{booktabs}
\usepackage{caption}
\usepackage{comment}

\usepackage{cleveref}
\usepackage{color}
\usepackage{enumitem}
\usepackage{float}
\usepackage{fancybox}
\usepackage{graphicx} 
\usepackage{xcolor}

\crefname{section}{Section}{Sections}
\crefname{subsection}{Section}{Sections}
\crefname{subsubsection}{Section}{Sections}

\renewcommand\footnotetextcopyrightpermission[1]{}

\setcopyright{acmcopyright}
\copyrightyear{2026}
\acmYear{2026}
\acmDOI{XXXXXXX.XXXXXXX}

\acmConference[CHI '26]{Proceedings of the CHI Conference on Human Factors in Computing Systems}{April 13--17, 2026}{Barcelona, Spain}
\acmBooktitle{Proceedings of the CHI Conference on Human Factors in Computing Systems (CHI '26), April 13--17, 2026, Barcelona, Spain}

\AtBeginDocument{%
  }

\begin{document}

\title[AI Agents for Human-as-The-Unit Privacy Management]{From Fragmentation to Integration: Exploring the Design Space of AI Agents for Human-as-the-Unit Privacy Management}

\author{Eryue Xu}
\affiliation{
  \institution{Northeastern University}
  \city{Boston}
  \state{Massachusetts}
  \country{USA}}
\affiliation{
  \institution{University of Illinois Urbana-Champaign}
  \city{Champaign}
  \state{Illinois}
  \country{USA}}
\email{eryuexu2@illinois.edu}

  \author{Tianshi Li}
\affiliation{
  \institution{Northeastern University}
  \city{Boston}
  \state{Massachusetts}
  \country{USA}}
  \email{tia.li@northeastern.edu}

\renewcommand{\shortauthors}{Xu and Li}

\begin{abstract}

Managing one’s digital footprint is overwhelming, as it spans multiple platforms and involves countless context-dependent decisions. Recent advances in agentic AI offer ways forward by enabling holistic, contextual privacy-enhancing solutions. Building on this potential, we adopted a “human-as-the-unit” perspective and investigated users’ cross-context privacy challenges through 12 semi-structured interviews. Results reveal that people rely on ad hoc manual strategies while lacking comprehensive privacy controls, highlighting nine privacy-management challenges across applications, temporal contexts, and relationships. To explore solutions, we generated nine AI agent concepts and evaluated them via a speed-dating survey with 116 US participants. The three highest-ranked concepts were all post-sharing management tools with half or full agent autonomy, with users expressing greater trust in AI accuracy than in their own efforts. Our findings highlight a promising design space where users see AI agents bridging the fragments in privacy management, particularly through automated, comprehensive post-sharing remediation of users’ digital footprints.

\end{abstract}



\begin{teaserfigure}
\includegraphics[width=\textwidth]{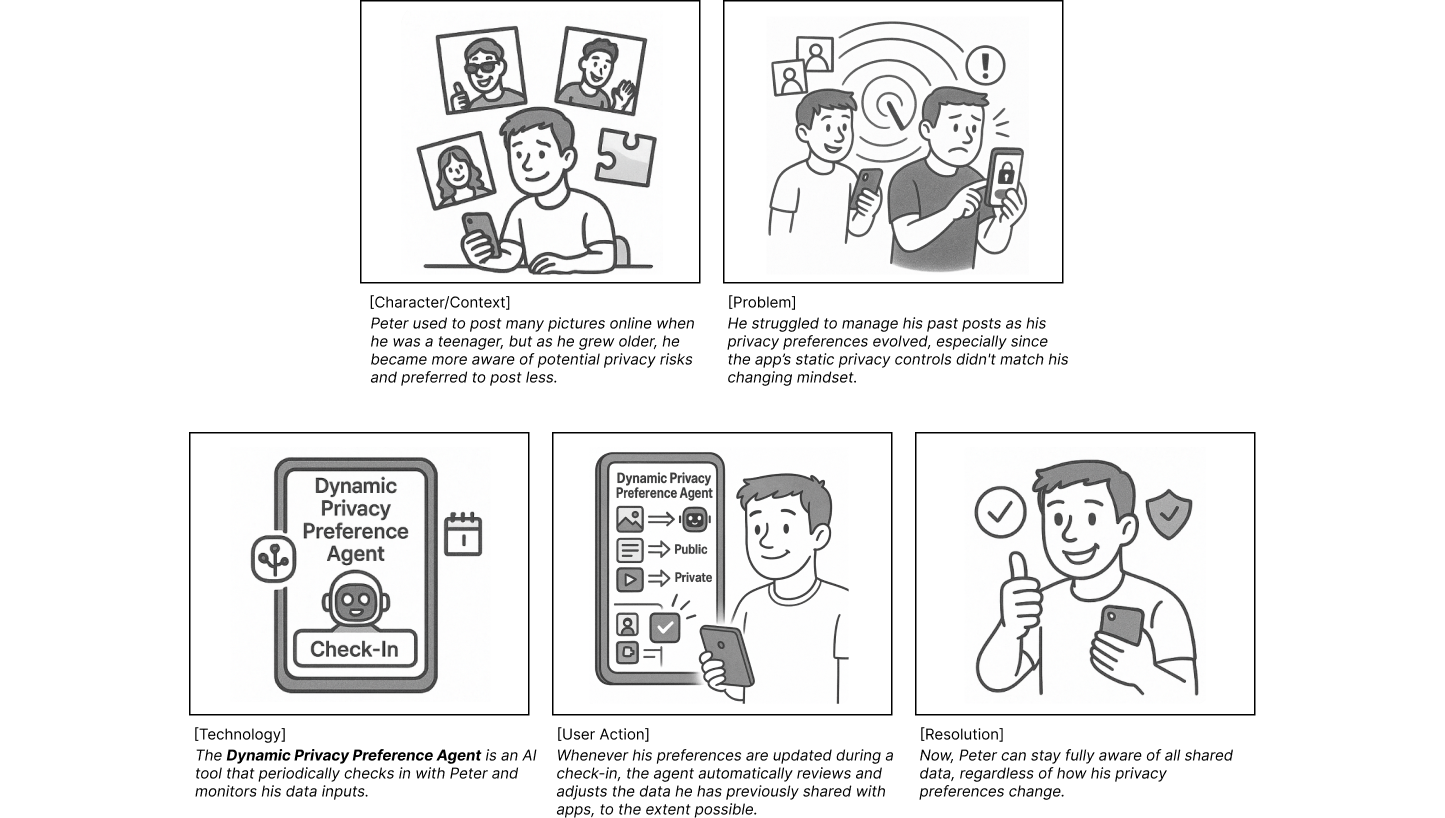}
\caption{Storyboard that illustrates the design idea of having a ``Dynamic Privacy Preference Agent'' to help user manage their past data shared history and align them with the user's most up-to-date privacy preference. This demonstrates one of the AI agent design concepts ranked highly by users among the human-as-the-unit privacy management proposals.}
\Description{A five-panel storyboard illustrating the user journey for the "Dynamic Privacy Preference Agent" design concept. The storyboard flows through the following stages: - Character/Context: Peter, who used to post many pictures as a teenager, becomes more aware of privacy risks as he grows older and prefers to post less. - Problem: Peter struggles to manage his past posts because the app's static privacy controls do not match his evolving privacy mindset. - Technology: The "Dynamic Privacy Preference Agent" is introduced as an AI tool on a mobile device that periodically monitors data inputs and checks in with Peter. - User Action: When Peter's preferences update during a check-in, the agent automatically reviews and adjusts his historically shared data (such as photos and videos) to align with his new preferences. - Resolution: Peter is shown giving a thumbs-up, satisfied that he can stay aware of all shared data regardless of how his privacy preferences change over time.}
\label{fig:Dynamic Privacy Preference Agent}
\end{teaserfigure}

\maketitle

\section{Introduction}

Continually, people describe their digital lives through a vocabulary of vulnerability: ``exposed,'' ``helpless,'' and ``overwhelmed,'' voicing the anxieties of privacy in a world where every trace lingers.
Researchers have long studied these phenomena under related constructs, including privacy fatigue~\cite{choi2018role,keith2014privacy}, digital resignation~\cite{draper2019corporate,draper2024privacy}, privacy cynicism~\cite{van2024privacy,hoffmann2016privacy}, and learned helplessness in privacy management~\cite{cai2024empirical, cho2022privacy}.
Various technologies and policies, such as platform-specific privacy controls,
have been created to protect people's privacy~\cite{apple2020privacy, android2023permissions,utz2019informed}, yet many people still sense that their privacy is at a vulnerable state~\cite{choi2018role, kang2015my,pappas2013assessing}.

Consider someone who once freely shared personal moments across platforms like Instagram and Facebook, accumulating hundreds of tagged photos. Over time, their preferences change and they want to hide some history from the public (Figure~\ref{fig:Dynamic Privacy Preference Agent}). They now face the daunting task of reviewing thousands of posts across multiple platforms---an overwhelming burden that existing privacy tools cannot adequately address.

This example is one of the many that reveal how current usable privacy controls fall short of people’s real-life needs. These shortcomings stem from two layers that could be investigated for more effective solutions. First, the dominant user-facing privacy research examines user perceptions within bounded contexts, such as specific applications (e.g., Amazon Alexa, ChatGPT, WhatsApp~\cite{lau2018alexa, 10.1145/3706598.3713701, munyendo2024you}) or technical domains including mobile, IoT, extended reality, and AI~\cite{degirmenci2020mobile, feng2021design,apthorpe2018discovering,hadan2024privacy,10.1145/3706598.3713701}. While these studies provide actionable insights to emerging privacy threats caused by new technologies, they overlook how individuals experience privacy as part of an ongoing, integrated digital life. In practice, privacy perceptions and behaviors transcend application, platform, or technology boundaries: people accomplish tasks across interdependent tools rather than within isolated platforms~\cite{fernandes2016appstract,liang2024understanding}; they are less motivated to protect information that they perceive as already disclosed elsewhere~\cite{kang2015my}; and they struggle to locate controls scattered across inconsistent interfaces~\cite{habib2020s, smullen2020best}. These dynamics point to a missing piece: the privacy-supportive tools that enable coherent management across individuals’ lived digital ecosystems.

Second, the fundamental granularity and complexity of people's privacy decision-making reveal a mismatch between the binary nature of most privacy controls (share/don't share, public/private), and the nuanced, context-dependent preferences users hold. Beyond the application- and technology-specific controls~\cite{10.1145/2556288.2557413, fang2010privacy, Seymour2020, 291152, 10.1145/3706598.3713701}, researchers have also introduced a set of system-level mechanisms like iOS privacy labels~\cite{apple2020privacy}, Android permissions systems~\cite{android2023permissions}, cookie consent banners~\cite{habib2022okay,schaub2016design,nouwens2020dark}, and policy summary and recap tools~\cite{10.1145/3340531.3417469}.
Although these controls help users manage a broader experience, they do not adequately account for the subtle and dynamic nature of disclosure contexts.
For instance, a user may disable location-sharing permissions across all social media apps, yet inadvertently disclose their whereabouts by posting details about their neighborhood, home environment, or school. Unlike data protected through system-level permissions, voluntary inputs---text, voice, image sharing, and UI interactions---pose distinct challenges due to their richness, granularity and reliance on self-censorship~\cite{wang2011regretted, zhang2024fair}.
This gap highlights a critical disconnect between broad, system-level privacy mechanisms and the fine-grained, everyday data disclosure decisions.

In this paper, we introduce the concept of \textit{human-as-the-unit privacy control}: mechanisms that center on people rather than apps or systems. This vision points toward a new generation of privacy controls that unify privacy management across contexts while supporting the nuanced, situational choices users make, empowering user agency over the complex privacy challenges in digital lives.

Recent advances in agentic AI present an unprecedented opportunity to realize human-as-the-unit privacy control. The state-of-the-art AI agents combine reasoning, planning, and tool use with the ability to interact with web services and graphical user interfaces~\cite{yao2023react, schick2023toolformer, wu2024autogen,openai2025operator}. These capabilities matter for human-as-the-unit privacy control because they enable cross-context operations to dissolve the fragmented silos of current privacy management, use persistent memory management to learn users' evolving privacy needs, and apply fine-grained decision logic to understand users' nuanced privacy preferences dynamically.
Early empirical and systems work has begun to explore this space. Recent studies have applied LLM agents to privacy-policy question answering and interactive policy summarization~\cite{sun2024empowering, chen2025clear, 10.1145/3706599.3719816}. Even adversarial demonstrations, showing that these agents can be used to extract sensitive data, serve as compelling proof of their technical potential~\cite{wang2025unveiling, kim2024llms, liu2024evaluating}. 
These developments demonstrate AI agents' potential to make human-as-the-unit privacy control feasible for the first time. 

Building on this potential, we adopt a design-exploration approach to map the space of AI-mediated, cross-boundary privacy management solutions. 
As an initial step toward addressing this broad gap, we choose mobile contexts as a substantial yet concrete domain since they are the primary point for everyday human-computer interaction, with 91\% of US adults owning a smart phone in 2025~\cite{pewresearch2024mobile}. 
These devices host apps spanning many facets of daily life, such as social networking, shopping, finance, health, and education, which together create cross-boundary privacy challenges~\cite{googleplay2025, appleappstore2025}.
Our investigation is driven by two research questions: 

\begin{description}
\item[RQ1] From a human-as-the-unit perspective, what cross-boundary privacy concerns do users encounter, and what strategies do they employ, if any, to address them?
\item[RQ2] What are potential opportunities for building AI-powered tools to support users' holistic privacy management needs?
\end{description}

Our research adopts a formative-to-summative flow, from exploration of user challenges to evaluation of design possibilities. We carried out our human-as-the-unit studies on a US sample. We investigated RQ1 by conducting 12 semi-structured interviews with participants who reported cross-boundary privacy concerns. From the interviews, we identified nine emergent human-as-the-unit privacy concerns that are undersupported by current technology and left to users’ manual efforts, encompassing cross-application boundaries, evolving temporal contexts, and granular interpersonal relationships. In parallel, we distilled two design factors. The first distinguishes between pre-sharing and post-sharing phases. The second captures levels of user agency, balancing control with convenience. Based on user needs and design factors, we developed nine AI agent concepts for privacy management, which are \textit{App Dictionary}, \textit{Digital Identity Manager}, \textit{Contextual Strategy Bot}, \textit{Summary Bot}, \textit{Dynamic Privacy Preference Agent}, \textit{History Sweeper}, \textit{Stimulated Privacy Community}, \textit{Post Central Manager}, and \textit{Auto Redactor}. Each design concept addresses one or more human-as-the-unit privacy needs and together they span the different variations of our design factors (\autoref{tab:storyboards_needs}, \autoref{tab:storyboards_design_factors}). 

To answer RQ2, we evaluated the nine privacy management agent concepts with 116 participants via a speed dating survey study, a well-established methodology for quickly examining a wide range of design ideas~\cite{davidoff2007rapidly,jin2022exploring}. Participants viewed storyboards that illustrated the context, problem, technology, user action, and resolution. 
They rated scenario relevance and concept effectiveness, and ranked their selections. Results showed that they are generally positive about both the identified needs and the perceived effectiveness of proposed solutions. Among the nine concepts, \textit{Digital Identity Manager} ranked highest, followed by  \textit{Dynamic Privacy Preference Agent} and \textit{History Sweeper}. The three concepts has an 80\%, 75\%, and 75\% chance to be preferred over the least ranked idea. Notably, the top three ranked ideas are all post-sharing management tools with half or full agent autonomy, indicating users' most-interested yet under-explored design opportunities. Our analysis surfaced the circumstances under which participants considered AI well-suited for privacy management, as well as the conditions they require for trusting and adopting it. We furthurly discussed the practicality and sociotechnical implications of AI agents for privacy management, offering a view of both the benefits and unforeseen risks.

To our knowledge, this is the first solution-focused study that examines privacy management from a human-as-the-unit perspective, moving beyond context-specific investigations. Our empirical findings reveal a critical insight about user preferences for privacy control: participants demonstrated stronger acceptance of AI agents managing their existing digital footprints compared to pre-sharing privacy protection. 
Their preference for post-hoc remediation informs future privacy tool designs by showing that users may be willing to cede control to intelligent systems when dealing with the complexity of their accumulated data traces. 
By demonstrating users' positive perceptions toward AI agents for cross-application, cross-time, and cross-network privacy management, our research opens new avenues for addressing the fundamental mismatch between users' mental models of privacy and the fragmented, platform-centric tools currently available to them.
\section{Related Work}

\subsection{User-facing Privacy Control}

Research on usable privacy controls has produced interventions that increase information transparency, expand user agency over privacy settings, and make privacy policies more accessible. Prior reviews have examined these efforts broadly. In this section, we organize these interventions by the contexts and boundaries they serve to map the current landscape and highlight existing gaps. 

\subsubsection{Application- and domain-specific solutions} 
A substantial literature develops user-facing tools that are scoped to the threat models and interaction affordances of particular app domains. Social networking applications have received extensive attention in early 2010s, as researchers investigated and deployed a variety of interventions for Facebook, including privacy nudges that prompt users to pause or revise posts, interfaces re-designs to make privacy settings more visible and actionable, and browser extensions that help users protect their profile information by generating obfuscated content~\cite{10.1145/2556288.2557413,10.5555/1387649.1387651, luo2009facecloak}.
In e-commerce, researchers have explored interfaces with improved findability and actionability, enabling users to more easily manage advertising and personalization settings~\cite{10.1145/3544548.3580773}.
In healthcare, work on patient-facing consent management and e-consent prototypes aims to empower patients with control over data sharing, while experimental systems seek to produce consumer-accessible compliance and risk reports for mobile health apps~\cite{10197085,Iafrate2016}.

\subsubsection{Emerging-technology specific solutions} For device ecosystems that blur the boundary between physical and networked contexts, researchers have produced tools that emphasize per-device visibility and local mediation. Work on perceptible assurance emphasizes tangible, physical affordances could improve users’ ability to verify microphone state and increase trust to the device~\cite{291152}. 
Similarly, smart-home projects introduced interactive frameworks that incorporate both digital and physical dashboards to visualize device locations and privacy assessments~\cite{windl2022saferhome}. Targeting to make assessments actionable, work on in-home assistants employed conversational explanations and simple firewall controls to help household members understand and manage their privacy options~\cite{Seymour2020}. Augmented-reality prototypes extend this design logic in two directions: visually mapping flows from physical devices to outgoing endpoints and letting users inspect and remediate leaks in situ~\cite{Cruz2023}, while adapting AR interface based on contextual privacy needs to prevent unintended information disclosure in shared spaces~\cite{10.1145/3706598.3713320}. These studies bridge the physical-digital privacy gap through tangible controls, spatial visualizations, and adaptive interfaces that make data flows perceptible and actionable at the device level.

\subsubsection{Platform-level and policy-oriented solutions} A third class of usable privacy interventions operates at platform or policy levels rather than within individual apps. Mobile platforms have taken divergent approaches: Both iOS and Android support runtime permission control over sensitive data and resources.
Android's open ecosystem enabled researchers to develop granular permission controls like sliders and context-embedded permissions~\cite{10.1145/3676519, 10.1109/SP.2012.24, 10.1145/2971648.2971693}.
Both platforms have also developed nutrition-label-like, succinct summary of the app's privacy practices, including Apple's App Privacy Details and Google's Data Safety Section, with researchers proposing expandable-grid designs to reduce information overload~\cite{apple2020privacy, 298912}.
Web-based privacy tools have similarly evolved from early frameworks like P3P~\cite{world2002platform} to sophisticated GDPR-era dashboards with layered consent options and cross-session preference management~\cite{schaub2017designing, nouwens2020dark}. Complementing these, policy-oriented tools make privacy policies themselves more accessible through automated assistants that navigate legal text~\cite{liu2019polisis, hong2019pribot}, browser extensions tracking policy changes~\cite{10.1145/3340531.3417469}, and systems highlighting non-compliant clauses~\cite{de2018claudette}. These platform-level interventions transform abstract privacy notices and choices into actionable guidance, enabling informed data-sharing decisions across digital ecosystems.

While the above interventions produce actionable privacy-enhancing supports within concrete contexts---such as social media, smart homes, or restricted platform use---they often narrow in too closely on specific problems to be broadly generalizable for individuals. Across these varied yet distinct problem spaces, privacy management challenges are highly specific---as users navigate multiple applications and devices daily, these solutions become fragmented. We shift focus to the design space of privacy supports that primarily address cross-boundary privacy protection, aiming to enable individuals to manage their privacy holistically across their entire digital lives.

\subsection{User Perceptions of Privacy Controls and Protective Behaviors}
\label{sec:related work user perceptions and bahaviros}

Researchers have conducted user studies to evaluate a variety of usable privacy controls they designed. Across these studies, while users consistently value transparency and fine-grained control, their actual experiences with privacy-enhancing tools remain mixed. Users are more willing to adopt systems that make data practices visible and provide granular options~\cite{10.1145/3676519, 10.1109/SP.2012.24}, yet too much complexity often backfires. For instance, privacy labels can overwhelm despite redesign attempts~\cite{zhang2022usable}, and nudges are frequently perceived as intrusive interruptions to natural interaction~\cite{10.1145/2556288.2557413, 11095630}. Research suggests that personalization may help mitigate this burden by aligning settings with individual preferences~\cite{windl2022saferhome}, though the trade-off between privacy and personalization remains debated~\cite{aguirre2016personalization}. At the same time, automation is welcomed when it reduces effort, such as through policy summarization~\cite{10.1145/3340531.3417469,freiberger2025prisme}. However, trust remains fragile: users doubt whether simplified summaries capture critical risks~\cite{gluck2016short, freiberger2025prisme}, and tangible indicators in smart devices offer limited reassurance when the data practices of underlying stakeholders remain opaque~\cite{291152}. These findings highlight that user perception of current usable privacy tools is shaped not only by their functional effectiveness, but also by broader concerns about trust and control that emerge across contexts.

In parallel, users do not passively rely on available tools, they actively develop their own strategies to manage privacy risks. Many practice self-censorship, strategically withholding information they deem sensitive~\cite{das2013self}, or segment identities through multiple accounts (e.g., “Finstas” or professional/personal profiles)~\cite{leavitt2015throwaway}. Others engage in obfuscation, from using throwaway accounts to falsifying personal details~\cite{sannon2018privacy}, enabling participation without full exposure. Progressive disclosure is another common tactic, with users initially sharing minimal information and gradually revealing more as they gain confidence~\cite{norberg2007privacy}. These behaviors reveal a pragmatic, sometimes effortful negotiation of risks, where users actively shape their own privacy boundaries.  

Overall, while research has developed privacy-enhancing tools to remediate shortcomings in specific apps, contexts, and systems, users, despite appreciating these features, report overload, distrust, or usability barriers that challenge adoption. In reality, they rely on folk-generated strategies in the field to manage their information disclosure choices. This gap underscores the need for privacy solutions that not only offer visibility or control, but also respect users’ natural strategies, and augment them with support that reduces effort and uncertainty rather than introducing additional burdens.  

\subsection{AI for Privacy Management and Analysis}

Researchers are increasingly developing privacy management tools with AI and LLM to help users navigate complex privacy decisions. A major portion of this emerging research focuses on AI-driven privacy policy analysis, addressing the fundamental challenge of policy understanding that users face. Recent work employs knowledge graphs for holistic policy representation~\cite{cui2023poligraph}, creates conversational privacy policy assessments browser extension~\cite{freiberger2025prisme}, demonstrates zero-shot analysis across previously unseen policies~\cite{tang2023policygpt}, and provides just-in-time risk identification which led to enhanced user awareness and improved privacy behaviors~\cite{chen2025clear}. Another trending direction is AI-assisted data obfuscation. Examples include smaller-LLM-powered ChatGPT user prompts redaction for enforcing data minimization~\cite{10.1145/3706598.3713701}, AI-powered image redaction to support blind and low vision people~\cite{zhang2024designing}, and intelligent risk identification with semantically preserving obfuscation for images~\cite{monteiro2024manipulate, monteiro2025imago}.

Beyond user-facing tools, LLMs are increasingly deployed in privacy engineering to address systemic challenges. This includes risk measurement, where LLMs identify and audit privacy vulnerabilities in large datasets~\cite{panda2025privacy, meisenbacher2025llm}; automated anonymization, leveraging contextual paraphrasing to preserve utility in sensitive text~\cite{yang2024robust, frikha2024incognitext}; and privacy-preserving data generation, where synthetic datasets replicate statistical properties without exposing individuals~\cite{bie2025generating,nahid2024safesynthdp}. These approaches illustrate LLMs’ essential role for scalable, infrastructure-level privacy protection.

While these AI-powered privacy management advances are promising, the field remains in its infancy as the AI era has just begun. Previous usable privacy research produced valuable but scattered context-specific solutions; meanwhile, LLM agents present a unique opportunity to unify these approaches as they advanced tool use, environmental sensing, and the ability to take contextual actions across different platforms, making them particularly well-suited to address the complex, cross-platform challenges inherent in human-as-the-unit privacy management~\cite{shavit2023practices,surla2025easy, yao2023react}. Therefore, the unexplored opportunity lies not just in technical feasibility but more in the breadth of design approaches that translate these capabilities into holistic user support. Our research investigates this design space---exploring how AI agents can be shaped to address individuals' comprehensive privacy needs, bridging the gap between what is technically possible and what users actually need for managing privacy across their digital lives.

\subsection{Privacy risk in AI}

The widespread adoption of AI systems introduces novel privacy risks that operate at a mix of technical and sociotechnical levels. AI systems exhibit inherent vulnerabilities in their architecture and training processes. Large language models have been shown to memorize or regenerate sensitive data, especially rare identifiers that persist in model parameters~\cite{carlini2021extracting, carlini2022quantifying}. This vulnerability is compounded by prompt-injection attacks, where crafted inputs can redirect a system’s decision or surface information never intended for disclosure~\cite{liu2023prompt, kim2024llms}. As models gain retrieval and tool-use abilities, their attack surfaces expand: adversarial systems such as RAG-Thief can systematically extract private information from retrieval-augmented knowledge bases through adaptive querying~\cite{jiang2024rag, zhang2025secret}.

To the human interaction level, the conversational and seemingly intelligent nature of AI systems ``seduces'' users to overshare personal information, amplifying inference risks via the centralized collection of conversational data~\cite{ma2025privacy, zhang2024fair, 10.5555/3766078.3766082}. 
While major technology companies have implemented safeguards and regulatory frameworks to govern data practices, the fundamental opacity of these systems persists~\cite{inan2023llama, li2023deepinception}. Human behaviors play another critical role when these powerful generative models democratize. The democratization lowers the barrier for malicious actors to conduct phishing, impersonation, and manipulation attacks with minimal expertise~\cite{hao2025spammers, heiding2023devising, li2026agentic}.

These multifaceted risks underscore why our work explores AI-powered privacy protection, recognizing that as humans increasingly integrate AI into all aspects of their lives, traditional privacy protection mechanisms alone cannot adequately address the scale and complexity of emerging threats. We envision agent-based approaches, which can provide holistic, contextual privacy management, operates at the same scale and sophistication as the AI systems generating these risks. Along with our vision, we speculated on and discussed how these vulnerabilities might impact our agent-based privacy management solutions. By offering a balanced view and proactive recommendations, we aim to help future scholars fully leverage the benefits of frontier AI agent technologies while cautiously mitigating potential side risks.

\section{Methodology}

Our study adopts a mixed-methods approach: a formative study of semi-structured interviews to uncover users’ privacy needs and data management practices (RQ1), and a summative study using a survey-based speed dating method~\cite{davidoff2007rapidly, jin2022exploring} to evaluate perceptions of AI-powered privacy solutions (RQ2). The studies were approved by the IRB at our institute. The interview protocol, survey design, and codebooks are attached in appendices.

\subsection{Phase 1: User Need Finding}

\subsubsection{Interviews}

We interviewed 12 participants who reported concerns about information disclosure across multiple applications. Participants first completed a screening survey in which they listed the applications they felt most concerned about. We imposed no restrictions on application genres or criteria, allowing us to capture representative sources of privacy concern beyond a single context. During 45-minute semi-structured interviews, we asked participants to reflect on (1) the types of information they had shared, (2) their specific cross-boundary privacy concerns, and (3) the strategies they had attempted to manage these concerns across applications. We piloted the interview protocol with four participants and refined our approach to probe cross-boundary privacy concerns along three dimensions: consistency, variation, and relatedness across applications. 

Our recruitment choice aligns with our study premise and prior work identifying cross-boundary privacy concerns as a meaningful entry point for examining users’ real-life privacy considerations. Focusing on participants with such concerns enabled us to elicit concrete scenarios, actions, rationales, and challenges that directly inform our design exploration. We intentionally did not solicit opinions about AI-agent-based privacy management to avoid prematurely constraining the design space, particularly given that AI agents are not yet widely adopted and may be difficult for participants to evaluate meaningfully within an interview setting.

To help participants appropriately examine and elaborate on concrete cases rather than abstract feelings, we used Amazon as an illustrative example, highlighting voluntary disclosures such as search keywords, customer service chat history, voice commands, reviews, and payment information. We then invited participants to articulate broader cross-boundary privacy concerns and explain the underlying goals of their actions, including the involvement of other parties when relevant. Throughout the interviews, participants described only the types of information disclosed rather than the actual content, ensuring that their privacy was protected.

\subsubsection{Analysis}
\label{sec:interview analysis}
We applied bottom-up thematic analysis, specifically affinity diagramming~\cite{braun2006using,harboe2015real}, for the interviews.  
We adopted an iterative process: the first researcher watched recordings of the interviews, reviewed the transcripts, created sticky notes for insightful quotes, and wrote down the thoughts on Figjam, an online collaborative whiteboard. The researcher completed thematic coding for the initial four interviews, after which the second researcher reviewed the Figjam whiteboard and provided feedback. The two researchers then met to collaboratively refine the themes and approach. Subsequently, the first researcher completed the analysis for the remaining 8 interviews while the second researcher closely reviewed all notes, confirming findings and suggesting edits for all participants.
The researchers then collaboratively conducted two rounds of clustering: (1) clustering user quotes that shared similar mid-level themes and identifying common themes, and (2) clustering related themes to form higher-level insights, which resulted in the nine key user needs and challenges reported in Section~\ref{sec:user needs} to answer RQ1. We also distilled two design factors from the nine key user needs and their ad-hoc strategies: (1) timing (pre-sharing, post-sharing) and (2) user agency (fully-autonomous, half-autonomous, user-controlled). Details are provided in Section~\ref{sec:design}.

\subsection{Phase 2: AI Solution Evaluation}

We developed nine AI agent design concepts for privacy management and illustrated each through a storyboard. We conducted a speed-dating survey~\cite{davidoff2007rapidly,jin2022exploring} with 116 participants to evaluate these concepts. This method is particularly effective in the early stages of design research, as it validates user needs with a sample and gathering feedback for the breadth of possible designs.

\subsubsection{Ideation}
Two coders, a PhD student and a CS professor (full positionality statement in Section~\ref{sec:positionality}), used FigJam to collaboratively develop nine storyboard ideas. Researchers began by displaying the nine key research insights from interviews, two design factors and a AI agent definition adopted from OpenAI~\cite{shavit2023practices} on the canvas to constrain the ideation space. 
This session generated 20 ideas in total. The researchers then converged, examining the ideas by mapping them to specific needs and outlining relevant design factors. From twenty brainstormed ideas, we consolidated into 9 ideas by merging overlapped ideas and discarding concepts that either did not leverage a necessary agentic AI capability or failed to explicitly solve one of the key research insights (the nine findings in ~\cref{sec:user needs}). For example, we discarded the idea of ``an AI educates users gradually about risks through interactive examples and privacy literacy moments" because it acted as a catch-all solution to increase user awareness rather than addressing a specific cross-boundary need.

\subsubsection{Storyboard Development}
To generate the corresponding storyboard for the final nine ideas, researchers authored a detailed description for each of the five frames, depicting the context, the problem, the proposed privacy protection concept, the user actions required, and the resolution.
To ensure visual consistency, we used GPT-4o-mini to generate the illustrations. 
Two researchers then reviewed all storyboards for quality assurance and conducted a comprehension check with six participants unfamiliar with the study. Each was presented with the nine storyboards and asked, ``To your understanding, what does this technology allow you to do?''
Across 54 responses, all participants correctly conveyed the intended meaning, with the exception of one participant who raised a concern about the technical feasibility of an AI illustration.
Such feasibility issues would not affect interpretation during the main survey, so we ensured our storyboard clearly conveyed the design concepts.

\subsubsection{Speed Dating Survey}
We used a survey format of speed dating~\cite{davidoff2007rapidly} established by prior research~\cite{jin2022exploring} to evaluate our nine concepts. Each participant viewed three randomly selected storyboards and rated each board: (1) scenario relevance to validate identified needs, and (2) perceived tool effectiveness to validate the concept. To avoid social desirability and confirmation bias from the storyboards' narration, we highlight in the opening of the survey that the design ideas are just concepts, and they are free to appreciate or criticize the idea solely based on their needs (\cref{appendix: survey}). 
We employed scenario-based questioning, asking participants to explain in open responses when they would use the effective tools or how they would modify the ineffective ones. This approach helps users situate themselves within the problem context and better envision the potential impact of a future solution. Their answers reveal the underlying rationales and decision factors influencing the adoption of AI-powered privacy solutions. The survey concluded with a ranking of their three viewed storyboards and an attention check requiring identification of the viewed storyboard numbers.
We disabled copy-paste functionality to discourage AI-generated responses.

\subsubsection{Analysis}
We applied the Plackett-Luce method~\cite{braun2006using} to generate a global ranking of the nine design concepts. For the qualitative answers, two researchers conducted the thematic analysis to identify factors influencing user acceptance of these AI privacy solutions.
One researcher first coded quotes and marked them with potential themes. The two researchers then met to discuss and refine the themes into two categories: factors that help users accept potential AI solutions and factors that need improvement to gain user acceptance.
We did not calculate the inter-rater reliability in the qualitative analysis because our goal is to identify emergent themes rather than seek agreement~\cite {mcdonald2019reliability}. The quotes presented in \cref{sec:AI pros} and \cref{sec:AI cons} were reviewed and examined by both researchers to be representative of the theme and collected data.

\subsection{Recruitment}
We recruited participants for both studies through Prolific. For the interview, we screened for US adult individuals who (1) expressed concerns about how their input data is handled across applications, (2) are fluent in English.
Out of 54 initial screening responses, 44 participants met our eligibility criteria.
We employed an iterative approach to recruitment and analysis to achieve theoretical saturation. We began by recruiting an initial batch of five participants, yielding four interviews. Two researchers thematically analyzed these findings (\cref{sec:interview analysis}). We then continued recruiting in batches of five while analyzing data concurrently. Recruitment ceased when both researchers agreed that subsequent interviews no longer produced substantive additions or revisions to the thematic structure. We reached theoretical saturation after 12 interviews.
To ensure diversity in technical familiarity, we sampled for educational background: 2 participants had a primarily technical background, 4 were non-technical, and 5 had a mixed or interdisciplinary background, and 1 reported as other (customer service). All participants were compensated 15 USD for their time.

For the survey, we stopped data collection once each storyboards were rated and compared at least 40 times, resulting in 144 responses. One researcher manually reviewed participants’ responses and filtered out unqualified entries (e.g., clear misunderstandings of all storyboards, low-effort answers, or meaningless answers). 116 participants passed the attention check and researcher's manual check for qualitative input quality. We decided not to collect more data based on established heuristics for choice modeling. Our study design (9 options, with 3 ranked per participant) results in the estimation of $8$ parameters ($K-1$) by the Plackett-Luce model. Our sample size provides approximately 15 participants for each parameter estimated, which is above the minimum of 10 participant-per-parameter heuristic recommended for stable model estimation~\cite{rao2003regression, bentler1987practical, hair2009multivariate}. 
Each participant was compensated with 4 USD.

\subsection{Positionality}
\label{sec:positionality}
The research team includes a doctoral student in information science and a computer science professor. The doctoral student has published work in usable privacy, security, and AI, and holds degrees in human-computer interaction. They are formally trained in qualitative coding and have developed strong expertise in AI and privacy. The computer science professor has an extensive research portfolio spanning usable security and privacy, human-centered AI, LLM agent, and AI privacy. Their experiences positioned the team to well analyze insights from studies, propose and critically assess AI agent designs for privacy management.

\section{Users' Privacy Protection Challenges from a Human-as-the-Unit Perspective (RQ1)}
\label{sec:user needs}

Our research shows that users' privacy concerns emerge from their integrated digital lives, rather than isolated use cases.
It may span multiple applications, evolve over time, and be shaped by dynamic interpersonal privacy boundaries.

\subsection{Challenges across the application}

\subsubsection{Information flow transparency}
\label{sec:flow transparency}
Participants expressed a universal concern regarding insufficient visibility into how their data is disseminated after sharing on an application. This lack of transparency undermines user agency, as participants originally granted one application access but observed subsequent cross-application sharing without anticipation or consent.
P6 specifically articulated concerns about data sharing with unknown third parties, explaining: \textit{``because you're dealing with third parties, you're dealing with different bidding, with different suppliers.''} The same participant also expressed frustration about interpersonal data sharing beyond their control, noting: \textit{``Even my friend would come into Facebook and download my photos and my profile picture without my consent.''} 
This dual concern shows that transparency challenges extend beyond institutions to interpersonal sharing.

Multiple participants (P5, P8, P11, P12) reported suspected data communication between applications owned by the same parent company. P12 described feeling \textit{``very uneasy about these three platforms''} when referring to Facebook, Instagram, and WhatsApp---all owned by Meta. Participants frequently used the term \textit{``manipulated''} to describe the consequences of opaque information flows, indicating a sense of vulnerability after voluntary disclosure.

\subsubsection{Inadequate privacy risk assessment}
Participants reported an absence of clear, accessible methods for understanding privacy risks or comparing policies across apps. Users often became aware of privacy vulnerabilities only after experiencing negative consequences, when remediation was less possible. 

Participants consistently cited unreadable privacy policy agreements as a primary barrier to informed decision-making. P3 and P11 emphasized this issue spans \textit{``many of these companies''}(P11) and expressed frustration about the lack of transparency.
Layered on top of these unreadable agreements, participants also attempted to develop their own comparative privacy assessment strategies across applications. For instance, P3 questioned a new finance app they were considering: \textit{``why do they collect signature while others do not need to?''} Yet participants could only pause usage, lacking tools or resources to understand the broader privacy landscape or systematically resolve concerns. Participants also relied on informal information-gathering strategies. P2 avoided using Temu and Shein based on privacy concerns, P4 expressed caution toward a particular bank, and P12 learned about Pinterest's problematic behaviors---all through scattered channels, including word-of-mouth recommendations and media coverage of privacy scandals.

\subsubsection{Identification protection challenges}
Participants sought selective anonymity while providing necessary functional disclosure, but consistently underestimated their identifiability via combined quasi-identifiers across platforms. P12 illustrated this vulnerability through a detailed example of cross-platform identification from Google, Strava (a fitness tracking app where users share real-name profiles and location data), and a dating app: \textit{``Let's say someone(on the dating app) wants to find me, they Google my name and Strava account pops up on Google."}, though she had not provided her full name on the dating app.
This shows how identifiable information voluntarily shared for one specific purpose can be a privacy vulnerability when accessed by unintended parties across different platforms.

Participants also faced barriers to implementing desired anonymity strategies. P3 and P4 expressed similar frustration: \textit{``If I can use a pseudonym, I'd rather not have any of this stuff associated with my real name. But unfortunately I cannot.''} and \textit{``If I can do pseudonym name on Uber, I will do.''} In response to platform limitations, multiple participants (P4, P8, P11, P12) developed consistent strategies across applications, including using non-face profile pictures and adopting nicknames where possible.

These findings reveal three aspects of identification protection challenges: participants recognize cross-platform linking risks, lack methods to assess overall identifiability, and adopt overlapping protective strategies---highlighting opportunities for system-level tools to unify and strengthen these efforts.

\subsubsection{Misaligned expectations with various platform privacy design}

Participants aced inconsistent availability across applications and lacked visibility into privacy controls, creating challenges for making informed decisions and maintaining consistent protection strategies. In highly sensitive contexts, participants manually applied sophisticated protections when platforms lacked native controls. For example, P5 and P12 deliberately misaligned timestamps with location data when sending social media pictures to protect real-time location. However, P12 struggled with Strava, which required her real name and simultaneously shared exercise information with her network, forcing her to accept unwanted visibility without knowing alternative options.

Participants also lacked awareness of evolving privacy features. P12 described quitting Uber due to harassment by a driver who accessed her personal information and remained unaware that Uber had since improved passenger privacy. When considering returning, she had to ask drivers directly whether her address could still be seen. These findings reveal how users struggle to maintain coherent privacy strategies when platforms offer inconsistent capabilities and users remain unaware of privacy improvements across their digital ecosystem.

\subsection{Challenges across the temporal context}
\label{sec:across time}
\subsubsection{Digital footprint management overload}
The proliferation of applications and accumulated digital traces create an overwhelming management burden for participants. Without regular audits, which few systems facilitate, users remain overwhelmed, as noted by P12: \textit{``there are so many apps at this point... it's hard to be actively concerned about all of them.''}

When reviewing the apps, multiple participants became acutely aware of information they had shared, revealing the extent of their previously unrecognized digital footprint. P2 expressed surprise: \textit{``Oh, my gosh! She knows a lot about me. 
''} P12 similarly described her realization: \textit{``The only reason I thought about it is because I took part in your questionnaire...I saw all of my previous addresses. Some of them I even forgot about that I lived there, but they still have it.''}

This awareness prompted immediate action intentions. P12 committed to scheduling time for data cleanup
as she characterized the scope of her accumulated data as representing her \textit{``whole life history,''} emphasizing the massive threat accumulated over time.

These findings demonstrate that users often remain unconscious of their extensive digital footprints until prompted to reflect systematically, revealing significant gaps in ongoing footprint awareness and management capabilities.

\subsubsection{Evolving privacy preferences without historical management tools}
As users develop more sophisticated privacy mental models through experience, they often struggle to retroactively apply new preferences to prior disclosures due to limited historical content management capabilities.

Participants generally evolved from naive early sharing behaviors to more privacy-conscious approaches over time. P5 described this process: \textit{``I really didn't have any internet safety skills... I signed up when I was 13. I used to post more personal information when I was younger. Now I don't really post... I've hidden my privacy settings and made some of my posts unavailable to the public.''} This shift from open sharing to selective controls illustrates a common developmental trajectory as users gain digital experience.

Some participants experienced significant mental model shifts without clear catalysts. P11 said, \textit{``Over 10 years ago, I didn't want to have social media anymore... I felt inundated... how in the world did this company or spam even find me... So I started being a lot more private.''} These changes occurred organically rather than through proactive, systematic privacy management.

\begin{table*}[t]
  \centering
  \caption{Nine Design Ideas with Descriptions and Key Needs Addressed}
  \label{tab:storyboards_needs}
  \renewcommand{\arraystretch}{1.3}

  \begin{tabular}{|p{0.5\columnwidth}|p{0.97\columnwidth}|p{0.49\columnwidth}| }
    \hline
    \textbf{Design Idea} & \textbf{Description} & \textbf{Need Addressed} \\
    \hline

    \textit{App Dictionary} &
    Use simple language and diagrams to showcase privacy assessment for applications, crowdsourcing from policies and reviews from forums, etc. &
    Information flow transparency; Privacy risk assessment \\
    \hline

    \textit{Digital Identity Manager} &
    Present existing public account information and provide suggestions on how users can anonymize themselves during account registration to prevent profiling. &
    Identification protection \\
    \hline

    \textit{Contextual Strategy Bot} &
    Suggest appropriate privacy customizations based on content sensitivity, platform, and audience. &
    Platform privacy variation \\
    \hline

    \textit{Summary Bot} &
    Show a summary of input from all applications to users at a set time interval, to remind users of all their digital footprints. &
    Digital footprint management \\
    \hline

    \textit{Dynamic Privacy Preference Agent} &
    Learn user preferences over time and manage historically shared data according to the most updated privacy preferences. &
    Evolving privacy preference \\
    \hline

    \textit{History Sweeper} &
    Delete the input history at a user-set time interval. &
    Accumulated data management \\
    \hline

    \textit{Simulated Privacy Community} &
    Simulate privacy preferences of each user and allow learning from each other's preferences to help protect others' privacy. &
    Interpersonal privacy boundary \\
    \hline

    \textit{Post Central Manager} &
    Learn user preferences over content dissemination and automate the process. &
    Relationship-based sharing \\
    \hline

    \textit{Auto Redactor} &
    Help edit text and obfuscate pictures on all platforms to a privacy-preserving point. &
    Identification protection; \newline Interpersonal privacy boundary \\
    \hline
  \end{tabular}
\end{table*}

\subsubsection{Accumulated data management deficiencies}
\label{sec:accumulated data}
Participants expressed strong preferences for managing their historical data and implementing data expiration policies, yet encountered substantial structural limitations in existing systems for tracking and managing data across platforms over extended periods. P8 advocated for systematic data lifecycle management: \textit{``I would probably make the companies... delete your information after a certain time period where it expires... I know I'm aware that anything you post on the Internet is basically gonna be there forever.''} Participants also demonstrated particular concern about being vulnerable to manipulation when getting old with P8 elaborating on long-term implications: \textit{``You're young, but you'll be old someday, and so everything you've done and purchased or searched for is all being accumulated... maybe you have an illness where you need help.''} This accumulated data burden further fostered a psychological shift from initial hesitancy to learned helplessness, as noted by P5: \textit{``I think I definitely was hesitant at first when I started using ChatGPT just because it was new... But I feel like I've shared so much of my data at this point, I don't really have any hesitation.''}
Without systematic tool support, participants either undertook overwhelming manual deletion efforts or, more often, resigned themselves to data accumulation as an unavoidable aspect of modern digital life.

\subsection{Challenges across interpersonal relationships}

\subsubsection{Interpersonal privacy boundary complications}
Participants managed information that implicated others' privacy in addition to their own, yet current systems rarely facilitate this need for collaborative privacy management. P6 articulated this responsibility: \textit{``I have to respect my friend's privacy. Maybe they don't want me to post it to the wild.''} , with other participants extended this consideration to children's privacy (P7, P12).
Participants also recognized how platform features can indirectly violate privacy despite users' attempts at control: \textit{``Sometimes you go on to someone's page and you don't see any of the pictures posted. But then you go into tagged and you see the whole history of a person, basically. And they didn't consent to it.''} (P12) This highlights the complexity of managing others' consent when platform features can circumvent individual privacy settings. 

Conversely, participants expressed concerns about others compromising their own privacy through shared access: \textit{``whenever you log in (youtube) to someone's machine... this person will have a glimpse of what you've been doing.''} (P10) In response to these challenges, participants developed manual workarounds, including image composition, \textit{``I hold my child with their back to the picture''} (P12), and systematic cleanup practices, always returning to browsers to delete traceable information after using shared devices (P10).

\subsubsection{Granular relationship-based sharing challenges}
\label{sec:across relationship}
Participants desired to control content visibility based on human relationships rather than platform-defined categories.
P12 wanted a relationship-based content control without specifying platforms: \textit{``That could be anonymized for the people that I choose it to be deanonymized for... Like family and friends
''} This reflects users' desire for granular control based on trust levels and relationship intimacy rather than broad platform categories like ``friends'' or ``public.'' P5 iterated on the same mindset that: \textit{``I wish all social media apps would have the ability to fully completely block someone... (now) if you block someone on Instagram, there's ways that you can get around it,''} revealing user desire to block or allow a data recipient by the ``person,'' but not an ID in an application. 

The granular nature of users' privacy needs extended beyond data recipient categories to contextual considerations, including content sensitivity, timing, and emotional vulnerability. P5 illustrated this complexity through a personal example: \textit{``I posted that my dog died. I would want that to be private because I don't want (that)... if someone didn't like me, they could tell me that they were happy my dog was dead.''} This demonstrates how seemingly trivial content can require sophisticated privacy protections based on users' understanding of their social relationships and potential risks.

\begin{table}[!t]
  \caption{Design Ideas and the Corresponding Design Factors, from Pre-Sharing to Post-Sharing, User-Controlled to Fully-Autonomous}
  \label{tab:storyboards_design_factors}
  \centering
  \small
  \renewcommand{\arraystretch}{1.3}
  \begin{tabular}{|p{0.19\columnwidth}|p{0.34\columnwidth}|p{0.34\columnwidth}|}
    \hline
    & \textbf{Pre-Sharing} 
      & \textbf{Post-Sharing} \\ \hline

    \textbf{User-Controlled} 
      & \textit{Contextual Strategy Bot, App Dictionary}
      & \textit{Summary Bot} \\ \hline

    \textbf{Half-Autonomous} 
      & \textit{Auto Reactor, Simulated Privacy Community}
      & \textit{Digital Identity Manager} \\ \hline

    \textbf{Fully-Autonomous} 
      & \textit{Post Central Manager}
      & \textit{History Sweeper, Dynamic Privacy Preference Agent} \\ \hline
  \end{tabular}
\end{table}

\section{Potential opportunities for building AI-powered tools to support users' holistic privacy management needs (RQ2)}

\subsection{Design concepts}
\label{sec:design}
This section presents the two design factors synthesized from user needs and their current ad-hoc control strategies, along with the nine AI agent privacy protection concepts.

\subsubsection{Design factor 1: Timing}
Several participants became aware of privacy risks only after irreversible negative consequences, including strangers appearing at their homes and a ride-share drivershowed up at their residence (P6, 11, 12). These reactive discoveries underscore the need for proactive risk identification tools. 
However, pre-sharing solutions alone are insufficient because (1) users’ mental models evolve over time, and (2) informed consent at the point of sharing does not ensure appropriate data use across its lifecycle. Once information is disclosed, users often feel helpless. This prompted them to adopt post-sharing strategies such as manually deleting digital histories (P8, P10, P11, P12).

To address this tension, we define pre-sharing tools as preventative mechanisms that identify risks before disclosure, and post-sharing tools as remedial mechanisms that help users manage data after sharing. Our design concepts intentionally incorporate both.

\aptLtoX{\begin{figure}[t]
  \centering
    \includegraphics[width=\linewidth]{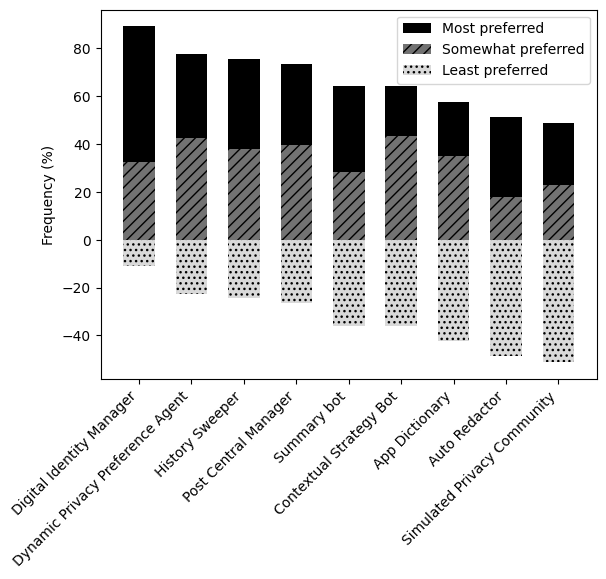}
    \caption{Rank distribution calculated by preferred order from Plackett-luce method~\cite{maystre2015fast}. From left to right, concepts are ranked in decreasing order of preference.}
    \label{fig:rank}
    \Description{A stacked bar chart showing the preference ranking of nine privacy design ideas. The x-axis lists the concepts ordered from most preferred on the left to least preferred on the right. The y-axis represents Frequency in percent. The bars are stacked with "Most preferred" (solid black) at the top, "Somewhat preferred" (diagonal stripes) in the middle, and "Least preferred" (dots) extending downward into negative values. "Digital Identity Manager" is the highest-ranked concept on the far left, showing the largest proportion of "Most preferred" votes. "Simulated Privacy Community" is the lowest-ranked concept on the far right.}
  \end{figure}
  \begin{figure}
    \centering
    \includegraphics[width=\linewidth]{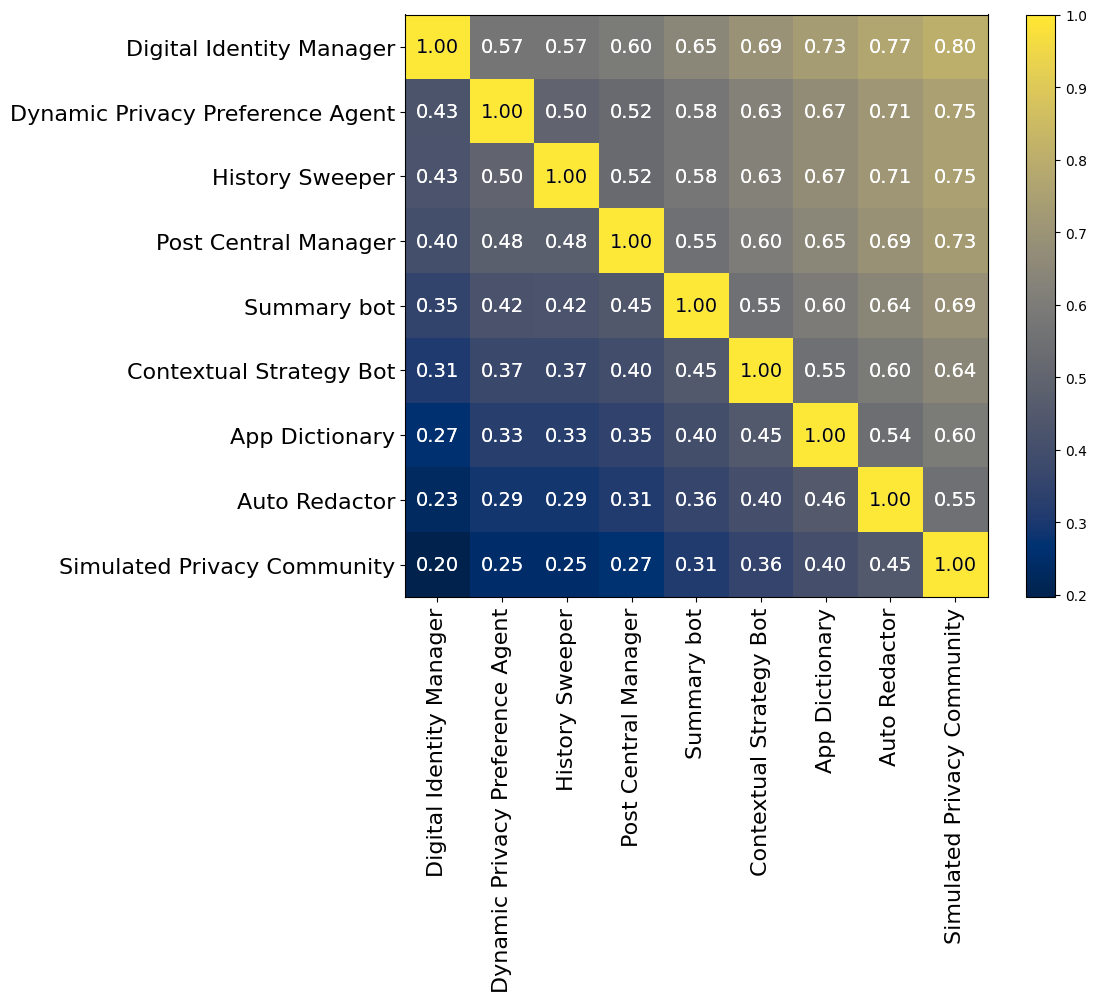}
    \caption{Probability distribution that a design idea (row) wins over another (column).}
    \label{fig:pair-wise}
    \Description{A heatmap matrix displaying the probability that a design idea in a row is preferred over a design idea in a column. The color scale ranges from dark blue (0.20 probability) to bright yellow (1.00 probability). The diagonal line is solid yellow (1.00). The "Digital Identity Manager" row (top) contains mostly light colors and high values (0.57 to 0.80), indicating it consistently wins against other concepts. Conversely, the "Simulated Privacy Community" row (bottom) contains dark blue cells with low values (0.20 to 0.45), indicating it rarely wins in pairwise comparisons.}
\end{figure}}{
\begin{figure*}[t]
  \centering
  \begin{minipage}[t]{1\columnwidth}
    \centering
    \includegraphics[width=\linewidth]{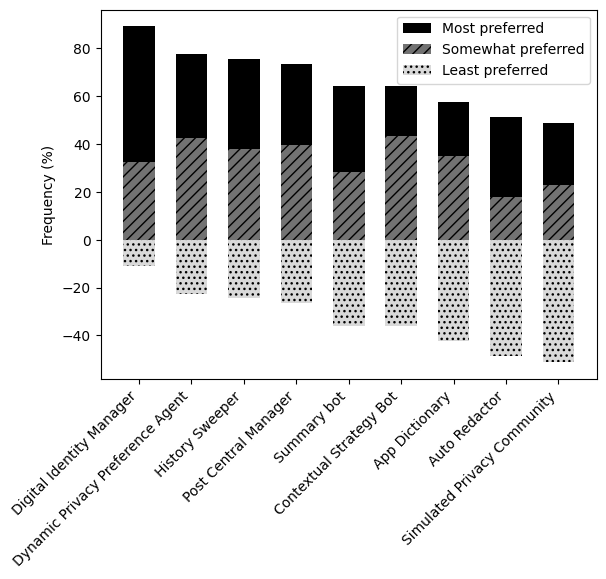}
    \caption{Rank distribution calculated by preferred order from Plackett-luce method~\cite{maystre2015fast}. From left to right, concepts are ranked in decreasing order of preference.}
    \label{fig:rank}
    \Description{A stacked bar chart showing the preference ranking of nine privacy design ideas. The x-axis lists the concepts ordered from most preferred on the left to least preferred on the right. The y-axis represents Frequency in percent. The bars are stacked with "Most preferred" (solid black) at the top, "Somewhat preferred" (diagonal stripes) in the middle, and "Least preferred" (dots) extending downward into negative values. "Digital Identity Manager" is the highest-ranked concept on the far left, showing the largest proportion of "Most preferred" votes. "Simulated Privacy Community" is the lowest-ranked concept on the far right.}
  \end{minipage}
  \hfill
  \begin{minipage}[t]{1\columnwidth}
    \centering
    \includegraphics[width=\linewidth]{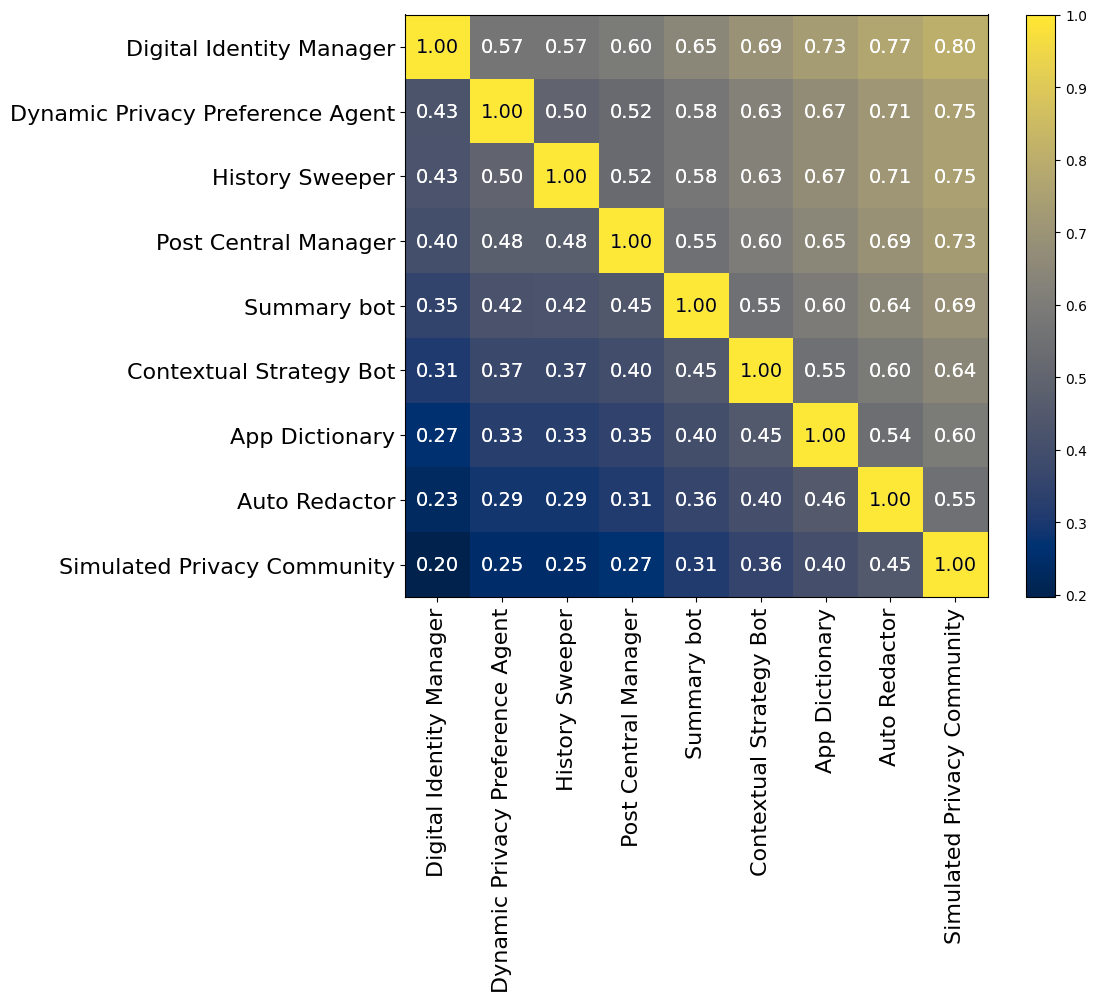}
    \caption{Probability distribution that a design idea (row) wins over another (column).}
    \label{fig:pair-wise}
    \Description{A heatmap matrix displaying the probability that a design idea in a row is preferred over a design idea in a column. The color scale ranges from dark blue (0.20 probability) to bright yellow (1.00 probability). The diagonal line is solid yellow (1.00). The "Digital Identity Manager" row (top) contains mostly light colors and high values (0.57 to 0.80), indicating it consistently wins against other concepts. Conversely, the "Simulated Privacy Community" row (bottom) contains dark blue cells with low values (0.20 to 0.45), indicating it rarely wins in pairwise comparisons.}
  \end{minipage}
\end{figure*}}

\subsubsection{Design factor 2: User Agency}
Participants described various ad-hoc information control methods, with a common pattern being the significant manual effort required for privacy management. For example, several participants discussed deliberately misaligning photographs and location data in their timelines or reported avoiding autofill features or maintaining separate accounts as privacy protection strategies (P5, 11, 12). These findings prompted us to investigate the optimal level of user agency---identifying which tasks participants prefer to have automated and which they prefer to control, potentially with AI support. Therefore, we identified three levels of user agency that vary across the nine design ideas: fully-autonomous, half-autonomous, and user-controlled. A fully-autonomous agent completes privacy management tasks without requiring user supervision. Half-autonomous agents complete privacy control tasks under user supervision, requiring user confirmation for key decisions. User-controlled solutions provide users with suggestions and knowledge, but require users to take the final privacy management actions.

\subsubsection{Final design concepts}
We consolidated nine design ideas based on the user needs identified in Section~\ref{sec:user needs}, spanning different design factor dimensions. Table~\ref{tab:storyboards_needs} summarizes these ideas with their descriptions and targeted needs, with Table~\ref{tab:storyboards_design_factors} categorizes them by design factors. Detailed storyboards are provided in Appendix~\ref{app:storyboards}.

\subsection{Speed Dating Quantitative Results}

\label{sec:quantitative}
We used the Plackett-Luce method~\cite{maystre2015fast} to calculate the worth of each idea globally and ranked them from most preferred to least preferred in Figure~\ref{fig:rank}. Among all options, \textit{Digital Identity Manager} ranked highest, with 21 of 37 participants ranking it as their most preferred tool out of the three options they evaluated.
\textit{Simulated Privacy Community} ranked lowest, with 20 of 39 participants ranking it as least preferred. Figure~\ref{fig:pair-wise} shows the probability distribution that an idea in each row wins over the idea in the corresponding column. For example, the probability that \textit{Digital Identity Manager} ranks higher than \textit{Simulated Privacy Community} is 80\%.

Table~\ref{tab:rank_design_factors} presents the ranked storyboards with their corresponding design factors and median scores for need validation and concept validation. The top three ranked design ideas are all \textit{post-sharing} privacy management controls. The first-ranked idea is \textit{half-autonomous}, followed by two \textit{fully autonomous} design ideas. 
Participants found the storyboard scenarios highly relatable and the design ideas generally effective (Figures~\ref{fig: needs} and~\ref{fig: concepts}). Most ideas received median ratings of 4 or 5 for both needs validation and tool effectiveness, indicating strong alignment with user expectations. \textit{Summary Bot} and \textit{Contextual Strategy Bot}, both user-controlled ideas, received the lowest median effectiveness score (3).

\begin{table*}[!t]
\centering
\caption{Ranked Ideas with Design Factors and the Median of Relatability \& Effectiveness (1 = the least, 5 = the most)}
\label{tab:rank_design_factors}
\begin{tabular}{c|c c c c c}
\toprule
 \textbf{Rank} &\textbf{ Design Idea}                     & \textbf{Timing} & \textbf{User Agency}      & \textbf{Relatability} & \textbf{Effectiveness} \\ \hline
 1 & \textit{Digital Identity Manager }          & Post   & Half-autonomous  & 5 & 4 \\ \hline
 2 & \textit{Dynamic Privacy Preference Agent}   & Post   & Fully-autonomous & 5 & 4 \\ \hline
 3 & \textit{History Sweeper}                    & Post   & Fully-autonomous & 5 & 4 \\ \hline
 4 & \textit{Post Central Manager}               & Pre    & Fully-autonomous & 5 & 4 \\ \hline
 5 & \textit{Summary Bot}                        & Post   & User-controlled  & 4 & 3 \\ \hline
 6 & \textit{Contextual Strategy Bot}            & Pre    & User-controlled  & 4 & 3 \\ \hline
 7 & \textit{App Dictionary}                     & Pre    & User-controlled  & 5 & 4 \\ \hline
 8 & \textit{Auto Redactor}                      & Pre    & Half-autonomous  & 4 & 4 \\ \hline
 9 & \textit{Simulated Privacy Community}        & Pre    & Half-autonomous  & 4 & 4 \\ 
\bottomrule
\end{tabular}
\end{table*}

\subsection{In what circumstances do users see AI as a good fit for addressing their privacy needs?}
\label{sec:AI pros}
Our analysis revealed that AI-powered privacy tools were perceived as most valuable when they addressed the challenges in users' current privacy management practices. 

\subsubsection{Adaptive Intelligence for Nuanced Preferences and Platform Variation.}
\label{sec:adaptive intelligence}
Participants believed AI-powered solutions excel at understanding nuanced privacy preferences that exceed the capabilities of standard application settings. 
S65 expressed needs for granular control beyond current application segments: \textit{``Posting professional content on LinkedIn, but wanting to exclude certain connections. Sharing personal updates on Facebook or Instagram, but only wanting specific friends or groups to see them.''} S2 illustrated the benefit of offloading nuanced preferences to AI agents: \textit{``If I wanted to upload photos of multiple people but wanted to respect their privacy... I wouldn't have to remember it each time,''} showing users expect AI to manage context-dependent preferences difficult to track manually.

Participants also valued AI's ability to handle platform variation and complex privacy information. S10 appreciated a bot that explains what privacy options mean, S42 expected to see tailored feedback on settings based on each site’s terms of service, and S113 requested content-level modification such as: \textit{``there should be an option to blank out children's faces with just their skin tone.''} All above examples illustrate a broad expectation for AI to translate complex intentions into precise protections across diverse contexts.

\subsubsection{Trust in AI's Accuracy Over Manual Processing.}
\label{sec:accuracy}
Participants felt that AI-powered privacy tools could manage their information more accurately than manual human processes. S6 articulated how AI's accuracy would increase their confidence in social sharing: \textit{``Whether I'm sharing memories with friends, posting highlights from a school project, or uploading content from a public event, I'd feel more confident knowing that everyone's privacy is being respected accurately.''}  Participants explained they found the \textit{Post Central Manager} helpful because it can: \textit{``minimize human errors,''} (S40) and prevent critical mistakes as it avoids \textit{``accidentally share it (a post) with the wrong network.''} (S55)
This trust suggests that users see AI agents not merely as convenient alternatives, but as superior solutions that can achieve levels of precision and consistency that exceed human capability in privacy management contexts.

\subsubsection{Enhanced Automation of Privacy Maintenance Workflows.}
\label{sec:automation}
Participants appreciated the potential for AI agent solutions to automate their information sharing processes, expressing enthusiasm for workflow benefits across three dimensions. They valued the prospect of reduced manual effort, particularly eliminating \textit{``extremely tiresome,'' ``manual deleting''} (S105) and handling complex social coordination by \textit{``automatically applying each person's privacy preferences without me having to ask them individually.''} (S6) Participants also emphasized the potential time savings through automated workflows. S96 envisioned the tool will automates posting across multiple social media platforms while preserving privacy for each audience, with 
S55 even directly quantified in mind that the solutions \textit{``make it 2x faster to organize and post content.''} Last, participants expected AI agents to updates workflows with changing circumstances: \textit{``The technology could be used to automatically adjust privacy settings on past posts as users' life circumstances or privacy preferences change, such as during career shifts or relationship changes.''} (S59)
These responses show participants' enthusiasm for AI-powered automation that would enhance their existing workflows and automate sophisticated, adaptive privacy management strategies.

\aptLtoX{\begin{figure}
    \centering
    \includegraphics[width=\linewidth]{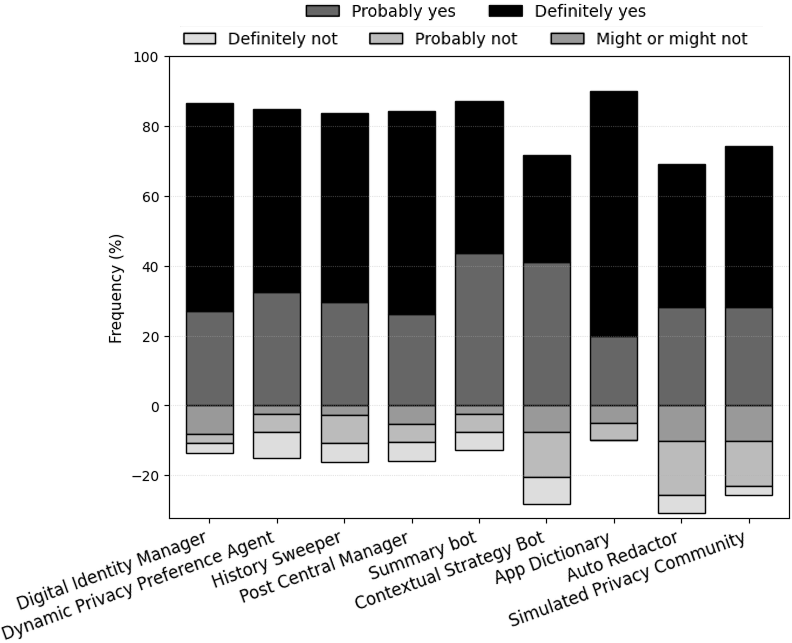}
    \caption{The distribution of answers to ``Could you relate to the concern?" Participants found App Dictionary's concerns most relatable, with Digital Identity Manager and Post Central Manager following in descending order. The left-to-right arrangement corresponds to the ranking shown in Figure~\ref{fig:rank}.}
    \label{fig: needs}
    \Description{A diverging stacked bar chart displaying user responses to the question, "Could you relate to the concern?" for nine design ideas. The y-axis shows Frequency percentages. Positive responses ("Definitely yes" in black, "Probably yes" in dark grey) extend upward from the zero line, while negative responses ("Definitely not" and "Probably not") extend downward. "App Dictionary" shows the strongest relatability with the largest "Definitely yes" segment. "Simulated Privacy Community" shows the highest frequency of negative responses.}
  \end{figure}
  \begin{figure}
    \centering
    \includegraphics[width=\linewidth]{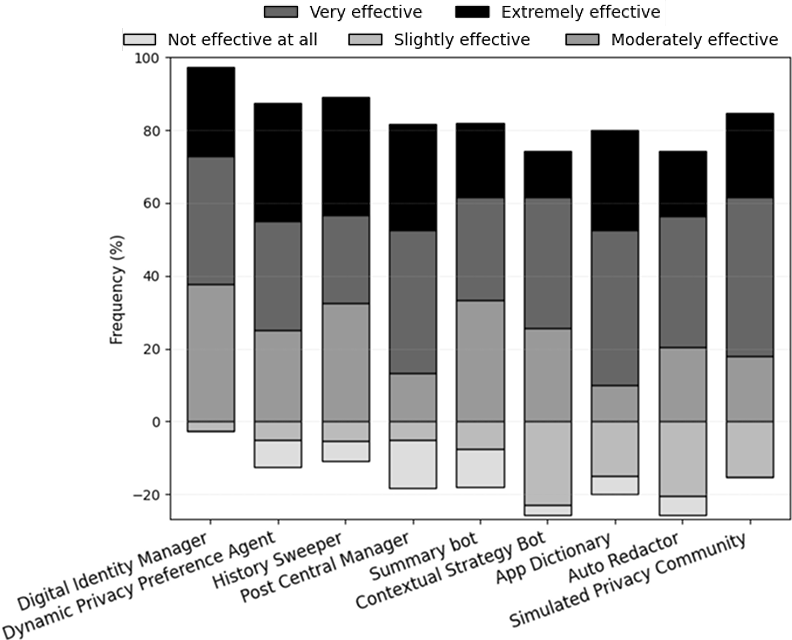}
    \caption{The distribution of answers to ``How effective does this solution address the concern?" Participants found Digital Identity Manager most effective, with History Sweeper and Dynamic Privacy Preference Agent following in descending order. The left-to-right arrangement corresponds to the ranking shown in Figure~\ref{fig:rank}.}
    \label{fig: concepts}
    \Description{A diverging stacked bar chart displaying user responses to the question, "How effective does this solution address the concern?" for nine design ideas. Positive ratings ("Extremely effective" in black, "Very effective" in dark grey) are stacked upwards, while negative ratings ("Not effective at all" and "Slightly effective") extend downwards. "Digital Identity Manager" (far left) appears as the most effective solution, having the tallest positive column. "App Dictionary" and "Simulated Privacy Community" show larger segments of "Not effective" responses compared to the other concepts. The left-to-right arrangement corresponds to the ranking shown in Figure 2.}
\end{figure}}{
\begin{figure*}[h]
  \centering
  \begin{minipage}[t]{1\columnwidth}
    \centering
    \includegraphics[width=\linewidth]{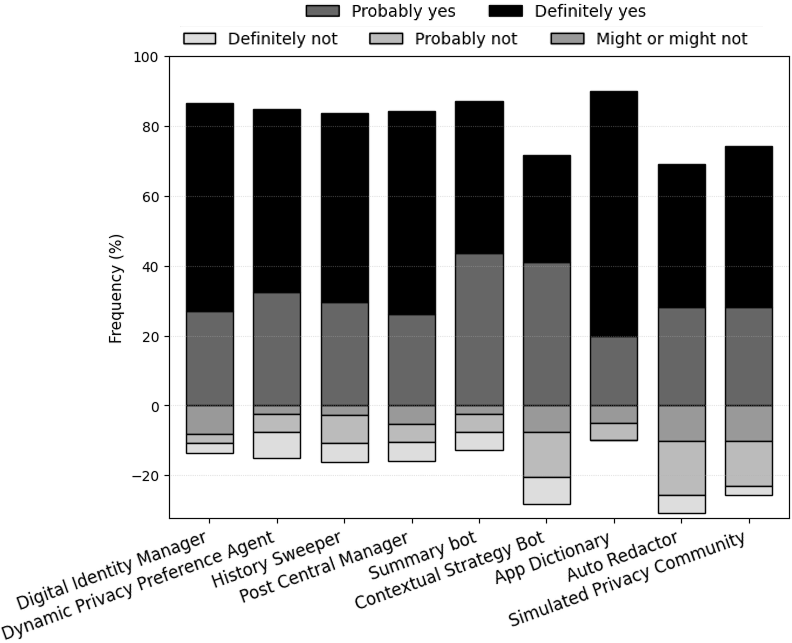}
    \caption{The distribution of answers to ``Could you relate to the concern?" Participants found App Dictionary's concerns most relatable, with Digital Identity Manager and Post Central Manager following in descending order. The left-to-right arrangement corresponds to the ranking shown in Figure~\ref{fig:rank}.}
    \label{fig: needs}
    \Description{A diverging stacked bar chart displaying user responses to the question, "Could you relate to the concern?" for nine design ideas. The y-axis shows Frequency percentages. Positive responses ("Definitely yes" in black, "Probably yes" in dark grey) extend upward from the zero line, while negative responses ("Definitely not" and "Probably not") extend downward. "App Dictionary" shows the strongest relatability with the largest "Definitely yes" segment. "Simulated Privacy Community" shows the highest frequency of negative responses.}
  \end{minipage}
  \hfill
  \begin{minipage}[t]{1\columnwidth}
    \centering
    \includegraphics[width=\linewidth]{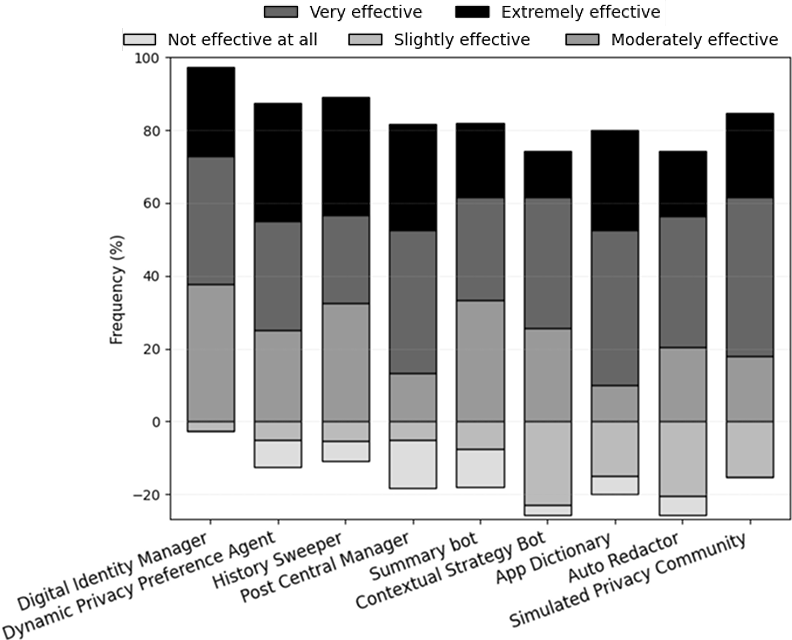}
    \caption{The distribution of answers to ``How effective does this solution address the concern?" Participants found Digital Identity Manager most effective, with History Sweeper and Dynamic Privacy Preference Agent following in descending order. The left-to-right arrangement corresponds to the ranking shown in Figure~\ref{fig:rank}.}
    \label{fig: concepts}
    \Description{A diverging stacked bar chart displaying user responses to the question, "How effective does this solution address the concern?" for nine design ideas. Positive ratings ("Extremely effective" in black, "Very effective" in dark grey) are stacked upwards, while negative ratings ("Not effective at all" and "Slightly effective") extend downwards. "Digital Identity Manager" (far left) appears as the most effective solution, having the tallest positive column. "App Dictionary" and "Simulated Privacy Community" show larger segments of "Not effective" responses compared to the other concepts. The left-to-right arrangement corresponds to the ranking shown in Figure 2.}
  \end{minipage}
\end{figure*}}

\subsubsection{As a Centralized and Comprehensive Hub.}
\label{sec:centralized}
Participants valued AI's ability to function as a centralized hub for privacy management, capable of coordinating diverse data across platforms. 
Participants appreciated AI's ability to centralize multi-platform privacy management. S37 explained: \textit{``I can see where I would use such an app if...I had photos, videos, and text posts on multiple social media platforms and wanted one app to go into those apps and make updates to my content rather than needing to go into each app myself.''}
This comprehensive approach could effectively address limitations in current privacy management and reflects users' holistic conceptualization of privacy. S81 further articulated: \textit{``If I could see a summary of all the emails, applications, CVs and photos I've sent out that would be a big help.''} In parallel, S74 and P37 highlighted participants’ concern for multiple content types, including visual, text, and multimedia, emphasizing the need for AI to manage diverse information comprehensively. These responses show participants view their information as an integrated whole rather than isolated silos, suggesting effective AI agents must support a comprehensive, multi-modal approach to privacy management.

\subsubsection{Proactive Privacy Management.}
\label{sec:proactive}
We found that the AI agent was expected to heighten users' human-as-the-unit privacy management experience by anticipating vulnerabilities and proactively managing potential risks. Participants like AI interventions that actively monitor and mitigate privacy threats before they escalate into significant problems, claiming \textit{``I would appreciate being told proactively if something risky might happen with my information.''} (S12) This proactive approach was especially valued because it shifted monitoring responsibility from users to AI agents, reducing the need for constant self-driven vigilance. One participant described the relief of having an agent handle oversight: \textit{``I use a lot of apps every day, for chatting, searching, shopping, and even health tracking, so it's easy to lose track of what I've shared and where. With a monthly summary, I'd be able to quickly review my activity across all apps, spot any potential privacy risks''} (S6). These examples highlight participants’ expectation that AI agents could shift privacy work from self-monitoring to situated response, allowing users to act on surfaced privacy insights rather than constantly generating them on their own.

\subsubsection{Reducing Cognitive Load and Easing Stress.}
\label{sec:cognitive load}
AI agents could effectively address privacy needs by significantly reducing cognitive load, enabling users to manage and reflect on their privacy practices without feeling overwhelmed. This reduction in cognitive burden directly alleviates user stress, providing a \textit{``peace of mind''} (S6, S15). One participant described the stressful nature of manually editing privacy settings, \textit{``I have done this before; editing is stressful.''} (S11) This highlights the broader mental pressures beyond specific privacy violations. Multiple participants shared similar emotional experiences. For example, S109 admitted feeling \textit{``guilty of not reading the textwalls that cover privacy policies''}, and praised the \textit{Summary Bot} solution, an AI-powered tool that summarizes lengthy, heterogeneous privacy information, as \textit{``super amazing and helpful.''} S105 also appraised our AI agent concept could \textit{``give me peace of mind knowing I'm more in control of my personal information''} and \textit{``save me worries.''} (S105) These responses demonstrate that effective AI agents address both practical privacy management challenges and the emotional burden of staying informed about privacy practices.

\subsection{In what circumstances do people not perceive the AI-powered tool useful, and what conditions must be met for them to consider it useful?}
\label{sec:AI cons}
Our analysis revealed that participants' skepticism toward AI-powered privacy tools stemmed from specific implementation doubts and trust concerns. Their willingness to adopt these tools was contingent on addressing several barriers.

\subsubsection{Independence of Interests and Hope for an Unbiased AI}
\label{sec:unbiased ai}
Participants expressed concerns about potential conflicts of interest in AI privacy tools, demanding independence from commercial entities that might compromise the tool's privacy-focused mission. S49 emphasized several key requirements, stating that AI-powered privacy-preserving technology must be \textit{``not released by any corporate, organizational, business, or personally involved individuals in any social media platform, i.e., it must be independent and not driven by an agenda beyond maximizing privacy.''} Similar concerns emerged in participants’ discussions of \textit{App Dictionary}. Reflecting on this design, S37 emphasized the need for AI systems to provide \textit{``unbiased recommendations''} when supporting privacy-related decision making. These accounts highlight participants’ view that trust in AI-mediated privacy support hinges on perceived neutrality and freedom from competing interests.

\subsubsection{Accuracy Concerns and Need for Human Supervision.}
Participants expressed particular concern about AI's ability to accurately understand and synthesize privacy-related information, especially regarding the \textit{App Dictionary} concept. S17 worried that \textit{``the AI would lose some words''} during processing, emphasizing that these words could be particularly important to users' understanding or could alter the meaning of critical privacy information. S70 expressed similar concerns with this idea and suggested implementing human supervision to oversee AI analysis. These concerns reveal a cautious outlook regarding AI's reliability in handling sensitive privacy information, where minor errors in interpretation could have significant consequences for users' privacy decisions. Participants' emphasis on human oversight suggests they view AI as a valuable assistant in privacy management, but not as a replacement for human judgment in critical privacy evaluations.

\subsubsection{AI Centralized Autonomy and User Agency}
\label{sec:cons:centralization}
While most participants responded positively to AI-powered privacy automation and valued centralized systems operating across applications and time spans, some expressed concern that concentrating control in a single AI agent could create a single point of failure. As S101 noted, \textit{``If the app gets hacked/compromised, all your important data is in one place.''} Participants also voiced broader unease about framing AI as a \textit{``catch-all term''} (S34) and about granting systems excessive autonomy. For example, S112 cautioned against entrusting an AI with full control over social media posts, stating, \textit{``This places too much power into the managing app's `hands.' I think it could potentially end up badly if it gets compromised.''} Instead of rejecting automation, participants emphasized the need for retained user agency through customization and oversight. Reflecting on the \textit{Auto Redactor} design, S89 suggested, \textit{``I'd want to be able to customize the type of information it censors,''} underscoring a desire to balance centralized AI support with meaningful human control in human-as-the-unit privacy management.
\section{Discussion}

\subsection{AI Agent for Post-sharing Interpersonal Privacy Controls}

Our speed dating results implicate a striking preference for post-sharing privacy management tools on \textit{interpersonal} privacy concerns. The top three ranked concepts all focus on addressing privacy concerns after data disclosure, with \textit{Digital Identity Manager} identifying existing searchable PII to help avoid cross-platform identity linkage, \textit{Dynamic Privacy Preference Agent} synchronizing previously entered content with current privacy preferences, and \textit{History Sweeper} automatically removing historical user inputs. This finding challenges the dominant paradigm in privacy research and design, which has focused predominantly on pre-sharing controls and preventive measures such as permission management~\cite{felt2011android, degirmenci2020mobile}, consent mechanisms~\cite{utz2019informed,degeling2018we}, 
data minimization~\cite{ganesh2024data,10.1145/3706598.3713701}, and data obfuscation~\cite{shokri2014privacy, monteiro2024manipulate, monteiro2025imago}. While these pre-sharing mechanisms are undoubtedly important, results from both studies (preference ranking from the summative study and the key user needs gathered in the formative study) demonstrate that users' desired privacy management extends far beyond the moment of initial disclosure. In \cref{sec:across time}, multiple participants highlight their desire to continue to manage the data life cycle, as the accumulated data represent their \textit{``whole life history''} (P12). P5's trajectory from naive sharing to privacy-conscious behavior on social media, exemplifies a common pattern where users' understanding of privacy risks matures through experience (\cref{sec:across relationship}). 

The emergence of AI systems amplifies these post-sharing privacy challenges in unprecedented ways. AI agents can retrieve documents from databases and dynamically inject this information into new contexts, elevating embedded data from low-level storage to high-visibility interfaces without users' awareness. 
Existing research on post-sharing data management has mainly focused on institutional privacy. For example, following the GDPR's provision of consumers' \textit{``right to be forgotten''}, ~\citet{habib2019empirical} has investigated the usability frictions on data deletion across commercial websites. ~\citet{peer2016impact} examined how reversibility notices influence users’ disclosure decisions when consenting to share personal information with companies. In these cases, platform controls and unified regulations could be useful for institutional privacy governance (\cref{sec:guideline}); however, our study identifies a persistent interpersonal privacy concerns of users’ lived experience that create gaps platform privacy controls and human-as-the-unit privacy concerns. In \cref{sec:user needs}, we have presented multiple user testimonies on worrying about specific interpersonal network gaining access to previously shared information. The users currently manage interpersonal privacy through fragmented and effortful strategies, including tweaking platform settings, creating separate accounts, or manually deleting posted content, lacking efficient tools (\cref{sec:related work user perceptions and bahaviros}). Aligned with our interview investigation, most of the top-half ranked ideas from stage two survey were designed for managing interpersonal cross-boundary privacy concerns (\cref{tab:rank_design_factors}).

We envision that this interpersonal focus makes AI agents for post-sharing management not only welcomed by user, but also practically verifiable. While confirming data deletion from corporate servers remains opaque---a fundamental challenge rooted in the asymmetric power and technical black-boxing of institutional data practices---interpersonal privacy management can be effectively achieved by AI agents operating at the GUI level, powered by their advanced capability to integrate tools and process natural languages~\cite{CrewAI2025, AutoGen2025}. For instance, the core capabilites for \textit{Digital Identity Manager}, the top ranked idea, to be feasible are interfacing with search tools (tool integration), interpreting retrieved content (natural language processing), and identifying potential inference risks arising from identifiers or combinations of quasi-identifiers~\cite{staab2024beyond} (natural language processing and reasoning). These agents can also manipulate the user-facing interface layer where interpersonal interactions occur, performing actions such as content deletion, visibility adjustments, and permission modifications through the same interfaces users themselves navigate. Moreover, we envision that AI agents can operate independently of individual platforms, echoing participants’ desire for an unbiased third party (\cref{sec:unbiased ai}). In this role, agents act as user-directed controls that more faithfully represent and adapt to evolving privacy preferences.

\subsection{User Agency on AI Cross-boundary Privacy Management}
We recognize that our proposed tools operate through different modalities of support, from teaching users about privacy risks, to presenting analyzed information for user action, to directly interfacing with applications on users' behalf. Understanding user preferences across these modalities illuminates how to balance automation with agency in Human-AI collaboration systems for privacy-critical contexts.
The ranking results challenge assumptions about user demand for control. The top three concept, \textit{Digital Identity Manager} (half-autonomous), \textit{Dynamic Privacy Preference Agent} (fully-autonomous), and \textit{History Sweeper} (fully-autonomous), all involve direct manipulation of their mobile apps, while purely informational tools ranked lower with median effectiveness scores of only 3 out of 5 (\cref{tab:rank_design_factors}). Participants articulated strong confidence in automated assistance (\textit{“minimize human errors,”} S40), which highlights the potential of bringing AI agent for privacy in work, as it suggests that existing privacy tools that focus primarily on information provision, such as policy summarization tools or privacy labels, while valuable for transparency, may insufficiently address users' actual privacy management burdens~\cite{kelley2009nutrition, apple_privacy_labels, 10.1145/3340531.3417469}. We see users expectation on agent automation as a practical projection because current workflow agents~\cite{CrewAI2025, AutoGen2025} have demonstrated advantages over purely manual workflows and current users increasingly view AI as an active collaborator~\cite{sundar2020rise}.

Besides their preference over automated solution, participants also valued the comprehensiveness of our ideation concepts as some ideas do not directly interact with their apps. Users particularly appreciate these agents ability to aggregate and present privacy information in new ways, such as generating summaries across multimedia sources, multiple applications, and longer time spans (S37, S74, S81). This finding supports our initial assumption that AI agents are particularly well-suited to addressing cross-boundary privacy concerns, a vision that is technically feasible given that recent multimodal large language models (e.g., GPT-5, Gemini, Deepseek-vl2) already demonstrate strong performance in summarizing heterogeneous information~\cite{team2023gemini, openai2025gpt5, wu2024deepseek}.
Participants described these aggregated summaries as a foundation for their proactive privacy management, e.g., \textit{“with a monthly summary, I’d be able to quickly review my activity across all apps, spot any potential privacy risks.”} (S6) and explicitly emphasized the need for human supervision (S17, S70, S89). These all reveal the importance to maintain meaningful user agency even when automation is available.

The above insights highlight users’ desire to remain cautious and actively engaged in managing their information disclosure in cross-boundary, privacy-critical contexts (\cref{sec:proactive}, \cref{sec:cognitive load}). This observation contrasts with long-standing findings on privacy fatigue~\cite{choi2018role, keith2014privacy}, digital resignation~\cite{draper2019corporate, draper2024privacy}, and privacy cynicism~\cite{van2024privacy, hoffmann2016privacy}. Instead, our results suggest that well-designed AI delegation may increase users’ overall agency by helping them stay aware of, and actively evaluate, what information they share---whether or not they rely on fully autonomous AI support.

Our approach recognizes that privacy-management solutions lie on a spectrum. At one end, AI agents can teach skills and provide information while preserving full user control. At the other end, AI agents can directly interface with applications to execute privacy-management tasks. However, users' enthusiasm for delegation coexists with concerns about vulnerabilities inherent in agentic AI systems, including risks of eroding privacy literacy and diminishing users’ situational awareness, ultimately weakening their ability to exercise independent privacy judgment. We expand on these risks in \cref{sec:challenges-risks} and propose corresponding design guidelines in \cref{sec:guideline}. Effective privacy agents must continuously navigate how different levels of autonomy distribute responsibility, transparency, and risk.

\subsection{AI Automation Bias and Risks}
\label{sec:challenges-risks}
While participants viewed AI agents as promising tools for alleviating privacy burdens, our findings also reveal important risks that emerge when these expectations intersect with the realities of current AI capabilities.
Given that LLM and agent technologies are fast-growing, so too is the landscape of privacy risks and biases they may introduce---risks that often require highly domain-specific expertise to fully anticipate and analyze. It is therefore challenging to expect general users to retain high awareness of these issues. Thus, beyond understanding what unmet privacy-management needs AI agents can address, our study also offers a window into potential user reactions to these powerful new technologies, which may include direct concerns and lack of trust, as well as overtrust and misplaced confidence. Below, we provide a critical analysis by aligning these user reactions with the risks revealed in cutting-edge research, which users may or may not foresee.

\paragraph{AI may lack nuanced contextual privacy reasoning.}
Participants frequently assumed that AI systems could deliver a level of precision and contextual sensitivity that remains unrealistic in today’s landscape (\cref{sec:accuracy}). Many of the top-ranked ideas, including \textit{Dynamic Privacy Preference Agent} and \textit{Digital Identity Manager}, required fine-grained inference of interpersonal boundaries, yet even state-of-the-art models struggle with nuanced relational privacy judgments and situational privacy norm reasoning, as noted in recent work on AI misalignment with social context and interpersonal norms~\cite{shao2024privacylens}. When such assumptions go unmet, the resulting expectation gap becomes fertile ground for over-trust. This over-trust often manifested as over-delegation. Several participants implicitly suggested giving the agent broad access so they could manage \textit{``everything''} (\cref{sec:centralized}), echoing well-documented patterns of automation bias and inflated confidence in AI systems~\cite{parasuraman2000model, romeo2025exploring}.

\paragraph{Single point of failure risk compounded by prompt injection vulnerabilities.}
Notably, some participants express concern about creating a ``single point of failure'' through centralized AI privacy management (S101, S112), highlighting a fear that one incorrect agent action could cascade across many linked accounts. This echos longstanding security tensions around centralized systems, from password managers~\cite{silver2014password,li2014emperor} to single sign-on services~\cite{sun2011makes}. Unlike password managers that store static credentials with clear boundaries, AI privacy agents elevates the risk by continuously accessing, interpreting, and acting upon a grand source of data across multiple platforms. In a recent Supabase Model Context Protocol (MCP) incident, the LLM IDE agent (Cursor) configured with elevated database privileges acted as a classic confused deputy: the agent interpreted untrusted user content (a support ticket) as executable instructions, queried sensitive tables, and returned secrets to the same channel~\cite{pomerium2025whenAIHasRoot}.
Therefore, we should be cautious to that privacy management agents need broad access to users' personal information and privilege to operate personal accounts on behalf of the users, while they may also be susceptible to these prompt injection attacks, leading to incorrect operations or even data theft.

\paragraph{Erosion of awareness and agency.}
Our findings point to broader sociotechnical risks rooted in long-term behavioral patterns even beyond these direct harms. Participants repeatedly conveyed the desire for the agent to monitor their privacy “proactively” (\cref{sec:proactive}) and reduce their cognitive burden (\cref{sec:cognitive load}), describing these features as offering \textit{``peace of mind''} (S6, S15). While these benefits are compelling, they also risk encouraging users to offload vigilance entirely---a phenomenon consistent with the erosion of situational awareness and “out-of-the-loop” effects documented in automation research~\cite{jiang2023situation}. Privacy research has similarly shown that persistent reliance on automated systems can diminish individuals’ sense of responsibility, understanding, and literacy around data practices, reinforcing patterns of digital resignation and reduced critical engagement~\cite{seberger2021empowering}.

Overall, these concerns illustrate a tension at the heart of AI-mediated privacy: the very features participants found most attractive, proactive management, reduced effort, and comprehensive surveillance across accounts, also create conditions in which users may over-trust, over-delegate, and lose the ability to detect when an AI agent acts incorrectly or oversteps. These risks motivate the need for the sociotechnical guidelines we outline next~
\cref{sec:guideline}, which seek to preserve user agency while enabling AI systems to operate safely within the complex realities of human privacy preferences.

\subsection{Towards Responsible AI Agents for Privacy}
\label{sec:guideline}
The risks identified in \cref{sec:challenges-risks} highlight the need for sociotechnical guidelines to support the safe and responsible deployment of AI privacy agents. A core design consideration is what data these systems collect, retain, or infer. This is the fundamental block for risks in misuse of data, inferences analysis and memory leakage in current literature~\cite{wang2025unveiling, liu2024evaluating, kim2024llms, staab2024beyond}. Many of the concepts in our study, for example, \textit{Dynamic Privacy Preference Agent} and \textit{Contextual Strategy Bot}, require access to users' personal data, activity traces, or platform metadata. However, as discussed in \cref{sec:challenges-risks}, granting broad or persistent permissions risks turning the agent into a single point of failure for privacy management. To mitigate this, future systems should adopt data minimization and purpose limitation, collect only what is required for a given task, and prevent silent accumulation of inferred attributes. Technical strategies such as on-device LLMs, constrained API permissions, encrypted or ephemeral memory states, and verifiable deletion workflows offer promising pathways for reducing inference risks \cite{10.1145/3706598.3713701, chen2024agentpoison}.

The needs and capabilities of interacting with the open worlds also increases the requirements of a layer of deterministic, verifiable control on top of the natural language commands that may be vague or be contaminated by malicious external content.
For example, when a ``History Sweeper'' cleans historical inputs that span multiple contexts, including those that accept data from external, untrusted sources (e.g., an email inbox that receives messages from others, or social media apps where other people can freely interact with the primary user), the agent might encounter confusing or intentionally misleading signals in this content, leading to incorrect deletion of data.
Future research and development of such agents should enhance their robustness against these adversarial outcomes by actively sanitizing commands, assessing risks, and implementing air gaps to isolate sensitive information and actions from adversarial interventions~\cite{10.1145/3658644.3690350}.

Yet technical controls alone cannot address the broader challenges surfaced in \cref{sec:challenges-risks} related to the erosion of human awareness and agency.
Responsible deployment will require organizational and policy-level governance that ensures transparency and accountability in agentic operations. Because AI agents can chain actions, invoke external APIs, and interact across multiple services, developers and institutions should establish auditable action logs, enforceable permission boundaries, and standardized disclosures about what data the system accessed and why. Such mechanisms help align users' mental models with system behavior, reducing the mismatch that often leads to over-trust and automation bias. Safe integration of AI privacy agents demands interfaces that balance automation with user agency. As shown in \cref{sec:proactive,sec:cognitive load}, participants valued proactive, cognitively offloading support, yet these same expectations can encourage over-delegation. To counter this tension, privacy agents should support adjustable autonomy, offer clear explanations of actions, and enable lightweight confirmation steps for sensitive decisions. These interaction patterns help align agent behavior with users’ nuanced and situational privacy preferences, including interpersonal boundaries, contextual sensitivity, and shifting emotional states, that AI systems cannot reliably infer on their own \cite{zhang2025towards, guo2025not, filipczuk2022automated}. 

Finally, while AI privacy agents offer one pathway for supporting privacy management, they are not a universal solution. As scholars of surveillance capitalism and cross-platform data flows have argued, many privacy harms originate not from individual decision-making but from structural incentives that encourage platforms to extract data and obscure privacy controls \cite{draper2019corporate}. AI agents may help individuals navigate this landscape, especially on mobile devices where boundary-crossing risks are most salient, but they cannot fully counteract platform-level practices designed to resist external oversight. This underscores the need to view agent-based solutions as complements, not substitutes, for stronger regulation, platform accountability, and cross-ecosystem enforcement. In combination, technical safeguards, governance mechanisms, and interaction designs can enable privacy agents to assist users without amplifying the vulnerabilities identified in \cref{sec:challenges-risks}, offering a more responsible pathway toward AI-mediated privacy in everyday contexts.

\subsection{Limitations}
This work is bounded by two scopes. First, we examine human-as-the-unit privacy management within mobile contexts, which limits the generalizability of our findings to other devices and cross-device ecosystems. Second, our studies primarily consider AI agents as a promising approach to addressing these cross-boundary privacy challenges, leaving a comprehensive analysis of non-AI technical approaches, organizational practices, or regulatory mechanisms beyond the scope of this work.

Our participants skew toward western and technologically proficient populations because we used a U.S.-based online recruitment platform. While this sampling strategy was appropriate for the current stage of the project and resource constraints, future work would benefit from engaging more diverse populations. Finally, Our findings may also reflect recall and confirmation bias inherent to self-reported data. 

\section{Conclusion}
Our research investigated users' cross-boundary privacy management needs and their perceptions of AI-powered solutions from a human-as-the-unit perspective. We identified significant gaps in how users currently manage information disclosure across their digital lives. When asked how relatable each scenario was to their own experiences, the speed dating participants gave consistently high ratings (median scores of 4-5 out of 5) for all proposed design concepts (\autoref{sec:quantitative}), demonstrating that the key privacy needs identified through in-depth interviews with 12 participants resonated broadly with a larger sample of 116 participants. Further, our speed dating participants identified several compelling reasons for wanting AI agent assistance with privacy management in~\autoref{sec:AI pros}, and these desires align closely with the current AI capabilities and near-term projections, revealing a major welcoming trajectory for AI-assisted privacy management. Finally, we discussed the implication of our study grounded with current privacy risk with AI agents, providing a balanced review for the AI agent for privacy tool builder. 

Our study reveals several critical directions for future research. Our findings highlight that post-sharing privacy management tools, particularly for interpersonal contexts, require sophisticated alignment mechanisms. Future work should investigate how AI agents can achieve dynamic preference learning while navigating the complex terrain of interpersonal privacy boundaries, understanding not just what to protect, but from whom and under what circumstances.
At the same time, the deployment of AI agents for privacy management introduces a fundamental paradox: these systems require extensive permissions and data access to protect user privacy effectively. Future research must address this tension between agent capability and security, exploring architectures that enable comprehensive privacy management without creating new vulnerabilities. 
\begin{acks}
We thank PEACH lab members for their thoughtful feedback on earlier drafts of this work. This work was supported by the National Science Foundation under Grant CNS-2426396.

\end{acks}

\bibliographystyle{ACM-Reference-Format}
\bibliography{bibliography}

\newpage
\appendix
\label{appendice}
\section{Storyboards}
\label{app:storyboards}

\section{Interview Screening Survey}
\label{app:screening}
Please complete the following form only if you are willing to participate in a 45-minute follow-up remote interview study with 15-dollar compensation. Your contribution would be highly appreciated!

\noindent\textbf{Q1. Open your mobile and briefly browse your installed applications. 
When using them, have you ever had privacy-related concerns about the information you’ve 
entered/input across different apps or platforms?} \\[2pt]
\emph{Consider various types of input, including search queries, comments on media posts, 
uploaded pictures, voice messages, any information that you have manually entered to the application counts.} \\[1pt]
$\bigcirc$ Yes, often. \; | \;
$\bigcirc$ Yes, occasionally. \; | \;
$\bigcirc$ No, not really.

\begin{figure*}[b!]
  \centering
  \setlength{\abovecaptionskip}{5pt} 
  \includegraphics[width=1\textwidth]{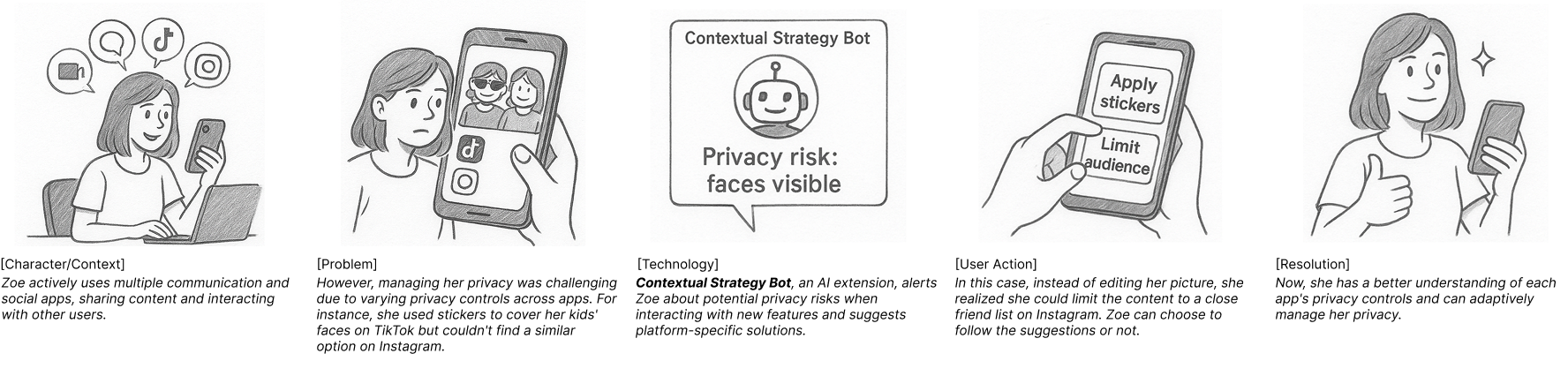}
  \caption{Storyboards 1: Contextual Strategy Bot }
\label{fig:Contextual Strategy Bot}
\end{figure*}
\begin{figure*}[b!]
  \centering
  \setlength{\abovecaptionskip}{5pt} 
  \includegraphics[width=1\textwidth]{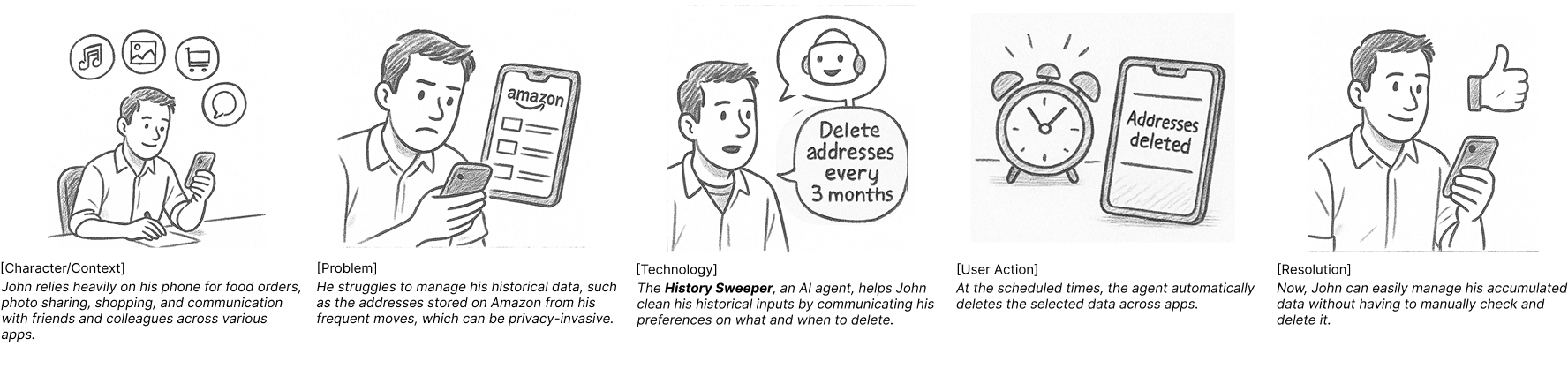}
  \caption{Storyboard 2: History Sweeper}
\label{fig:History Sweeper}
\end{figure*}
\begin{figure*}
  \centering
  \setlength{\abovecaptionskip}{5pt} 
  \includegraphics[width=1\textwidth]{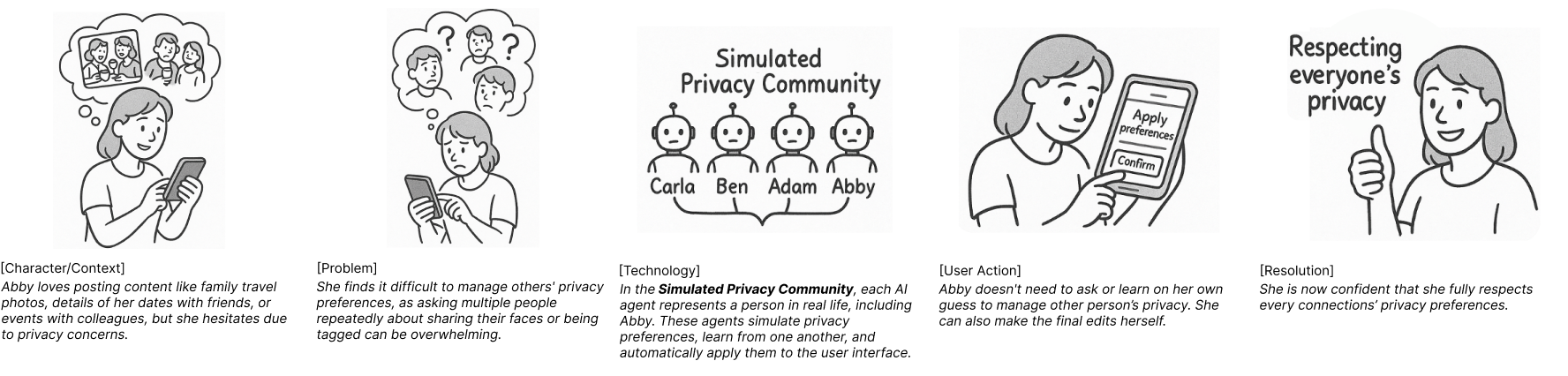}
  \caption{Storyboard 3: Simulated Privacy Community}
\label{fig: Simulated Privacy Community}
\end{figure*}
\begin{figure*}
  \centering
  \setlength{\abovecaptionskip}{5pt} 
  \includegraphics[width=1\textwidth]{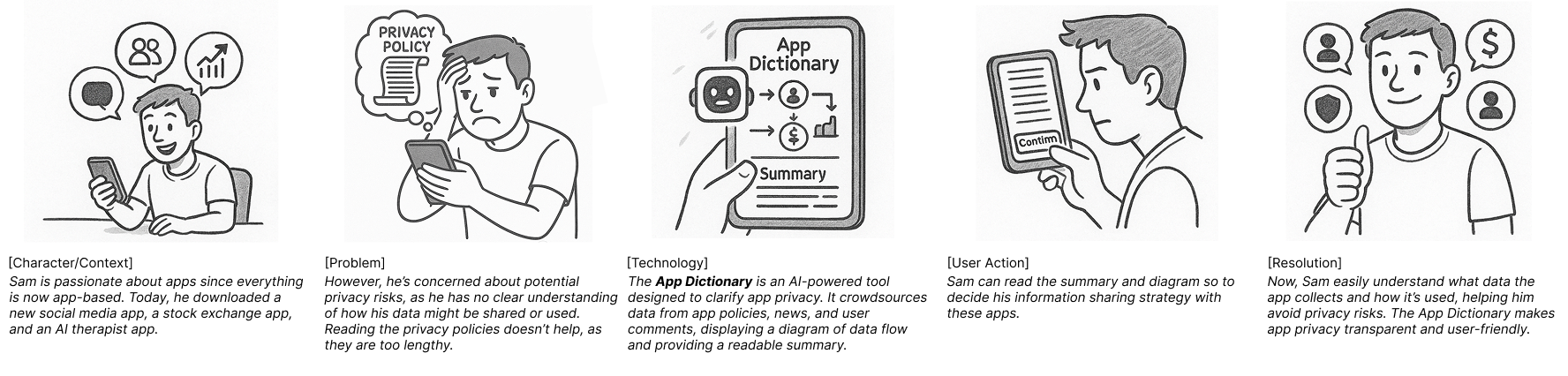}
  \caption{Storyboard 4: App Dictionary}
\label{fig:App Dictionary}
\end{figure*}
\begin{figure*}
  \centering
  \setlength{\abovecaptionskip}{5pt} 
  \includegraphics[width=1\textwidth]{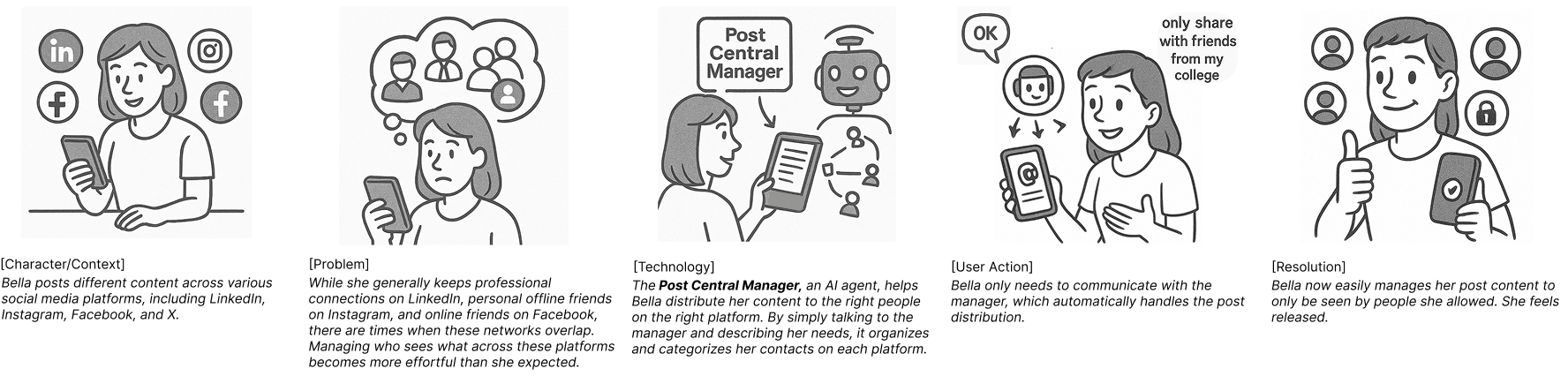}
  \caption{Storyboard 5: Post Central Manager}
\label{fig:Post Central Manager}
\end{figure*}
\begin{figure*}
  \centering
  \setlength{\abovecaptionskip}{5pt} 
  \includegraphics[width=1\textwidth]{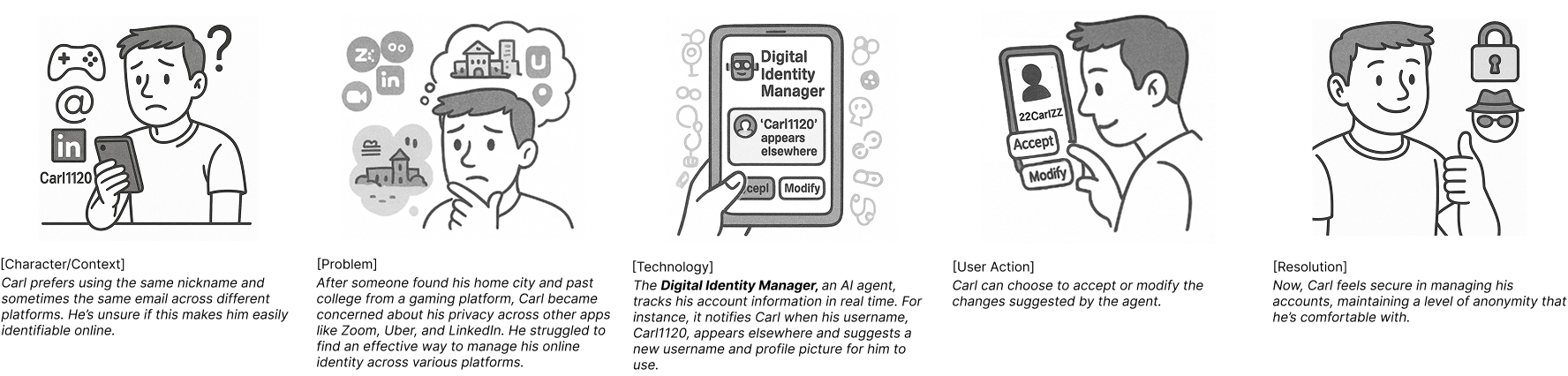}
  \caption{Storyboard 6: Digital Identity Manager}
\label{fig: Digital Identity Manager}
\end{figure*}
\begin{figure*}
  \centering
  \setlength{\abovecaptionskip}{5pt} 
  \includegraphics[width=1\textwidth]{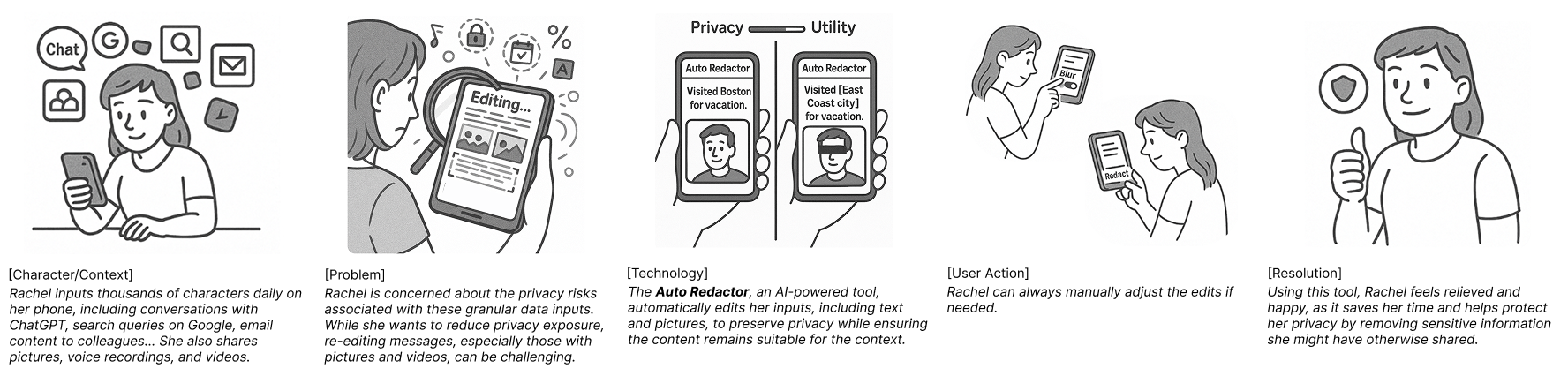}
  \caption{Storyboard 7: Auto Redactor}
\label{fig:Auto Redactor}
\end{figure*}
\begin{figure*}
  \centering
  \setlength{\abovecaptionskip}{5pt} 
  \includegraphics[width=1\textwidth]{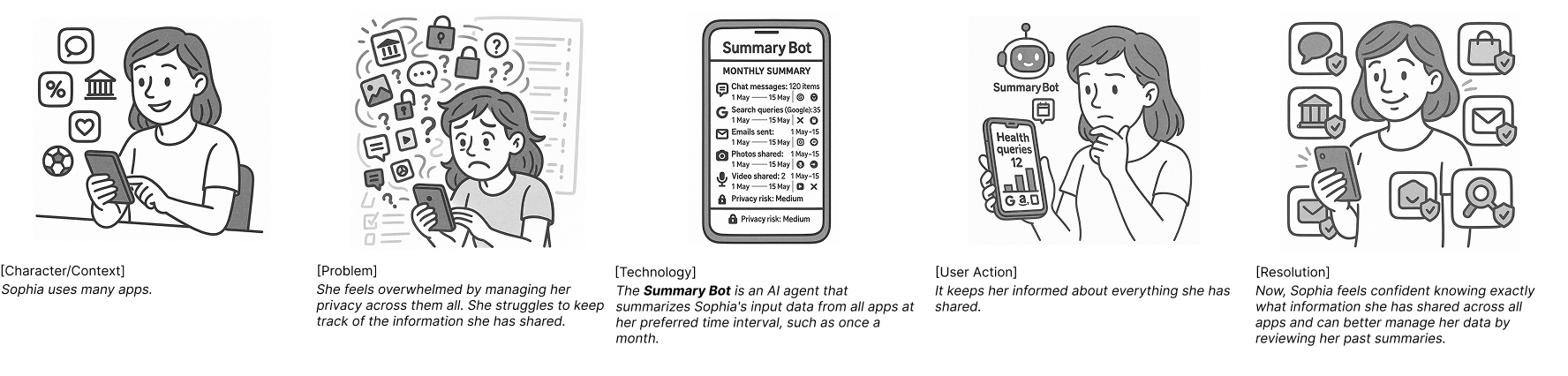}
  \caption{Storyboard 8: Summary Bot}
\label{fig:Summary Bot}
\end{figure*}

\noindent\textbf{Q2. What are eight apps that you are mostly concerned with your information input? You do not need to rank them in any order.} \\[2pt]
\emph{Consider the apps you use most frequently, as well as those where you enter particularly sensitive information, even if you use them less often. Hint: you can try open your mobile device screentime/battery usage applications list, and start filtering from there.} \\[2pt]
\noindent\textbf{Q3. What best describes your educational and professional background?}
$\bigcirc$ Primarily technical (e.g., computer science, engineering, IT) \;|\;
$\bigcirc$ Primarily non-technical (e.g., arts, humanities, social sciences) \;|\;
$\bigcirc$ Mixed (a combination of technical and non-technical education or work experience) \; | \;
$\bigcirc$ Other (please specify): \rule{3cm}{0.4pt}

\section{Interview Protocol}

Today’s study focuses on information you actively share with applications—text entries, posted pictures, voice messages, and other user-input data. Please review the 8 applications you listed in the survey. Open each app on your mobile device and examine the information you have input; let me know what you have shared with each app. To help you understand, here is an example: Suppose Amazon is listed. Open Amazon on your phone and note entries where you voluntarily shared data. By examining each entry, you can recall instances where you felt concerned.

For each application, ask:
\begin{itemize}[topsep=0.2ex, itemsep=0.2ex]
  \item Were there any instances where you thought twice before sharing? (e.g., afraid of inputting data to the company that owns the app; afraid of people—public, strangers online, someone you know—seeing the content; afraid of being identifiable; afraid of leakage of sensitive information)
  \item Are you concerned with any information you have input there? Why?
  \item If no concerns were mentioned: After review, what do you think the app knows about you?
  \item Have you adopted any ways to address your concern?
  \item How is this input strategy consistent or different from other apps?
\end{itemize}

When you take a step back and look at all the different ways you shared information across various apps:
\begin{itemize}[topsep=0.2ex, itemsep=0.2ex]
  \item \textbf{Consistency:} Do you notice any consistency in strategy and concerns over some apps? I noticed you mentioned similar concerns/controls over specific apps—why?
  \item \textbf{Variation:} What are some variations in your concerns and controls? You mentioned a specific concern/control using App 1; have you adopted it for App 2? Why or why not?
  \item \textbf{Relatedness:} Have you ever used two or more apps to get one thing done? How are they related? What information have you shared across these applications? Do you have any concerns with this flow?
\end{itemize}

Imagine you are a magician and can create anything without material or technological limitations. If asked to create one or more tools to address your concerns and facilitate your information-sharing process, what would they be like? This could target a specific group of apps, a single app, or all apps. It can be a concrete design idea or just a few desired capabilities. How will this idea help the specific cases you mentioned before?

\section{Survey Instrument}
\label{appendix: survey}
\textbf{Open Message}

\noindent Thanks for your willingness to contribute to our study!
\noindent We are interested in gathering your feedback on AI-driven app-input privacy management. In this 15-minute survey, you will be asked to evaluate several storyboards that illustrate different design ideas for managing app input privacy.
Note that these design ideas are just concepts—please assume that all are technically feasible and feel free to criticize or appreciate them solely regard your needs.

\noindent Please do not use LLM-based tool to generate your answer. We want to hear your real thoughts. It is OK to be drafty.
\noindent There is not a lot questions! So, we would really love you to take your time, think deeply, and explain your thoughts.
\noindent Your feedback will help us understand how these concepts resonate with users like you.

\noindent \textbf{Question for each storyboard}

\noindent \textbf{Q1} Could you relate to the person’s concerns shown in the storyboard? 
\(\bigcirc\) Definitely yes \(\bigcirc\) Probably yes \(\bigcirc\) Might or might not \(\bigcirc\) Probably not \(\bigcirc\) Definitely not

\noindent \textbf{Q2} How does this technology shown in the storyboard address the described concerns? 
\(\bigcirc\) Extremely effective \(\bigcirc\) Very effective \(\bigcirc\) Moderately effective \(\bigcirc\) Slightly effective \(\bigcirc\) Not effective at all

\noindent If the participant considers the technology effective in Q2:\\
\textbf{Q3} Could you describe some scenarios that you may use the technology shown in the storyboard?

\noindent If the participant does not consider the technology effective in Q2:\\
\textbf{Q4} How can this proposed solution be made more effective to address your need? For example, what would you like to change or add?

\noindent\textbf{Question for rank}

\noindent Please rank the technologies shown in the following scenarios
in order of effectiveness of addressing your privacy need in your
smart home, where 1 is most effective and 3 is least effective.

\section{Demographic}

\begin{table}[h!]
\centering
\begin{tabular}{l c}
\hline
\textbf{Demographics} & \textbf{Frequency} \\
\hline
\multicolumn{2}{l}{\textit{Age Range}} \\
18--29 & 3 \\
30--39 & 5 \\
40--49 & 3 \\
50--59 & 1 \\
60--69 & 0 \\
70--79 & 0 \\
\hline
\multicolumn{2}{l}{\textit{Gender}} \\
Female & 7 \\
Male & 5 \\
Prefer not to say & 0 \\
\hline
\multicolumn{2}{l}{\textit{Race/Ethnicity}} \\
White & 5 \\
Black & 2 \\
Asian & 4 \\
Mixed & 0 \\
Other & 1 \\
\hline
\end{tabular}
\caption{Demographics of interview participants (N=12).}
\vspace{-8pt}
\end{table}

\begin{table}
\centering
\begin{tabular}{l c}
\hline
\textbf{Demographics} & \textbf{Frequency (\%)} \\
\hline
\multicolumn{2}{l}{\textit{Age Range}} \\
18--29 & 27 (23.3\%) \\
30--39 & 37 (31.9\%) \\
40--49 & 23 (19.8\%) \\
50--59 & 16 (13.8\%) \\
60--69 & 10 (8.6\%) \\
70--79 & 3 (2.6\%) \\
\hline
\multicolumn{2}{l}{\textit{Gender}} \\
Female & 59 (50.9\%) \\
Male & 56 (48.3\%) \\
Prefer not to say & 1 (0.9\%) \\
\hline
\multicolumn{2}{l}{\textit{Race/Ethnicity}} \\
White & 75 (64.7\%) \\
Black & 19 (16.4\%) \\
Asian & 9 (7.8\%) \\
Mixed & 8 (6.9\%) \\
Other & 5 (4.3\%) \\
\hline
\end{tabular}
\caption{Demographics of survey participants (N=116).}
\end{table}

\section{Codebook}

\subsection{Interview study}
Each high-level finding is followed by categorized midlevel themes with numbers of quotes identified the in brackets.

\begin{itemize}
\item \textbf{Information flow transparency deficits (35)} 
  
    \textit{data ownership across multiple entities (3), concerns about the same parent company (16), apps are linked (9), unexpected data subjects (7)}
  \item \textbf{Inadequate privacy risk assessment mechanisms (48)}
  
    \textit{no regulated way to understand (8), concerns about the same parent company (16), no way to know until a consequence (14), unknown about the company (3), unknown about the regulation (2), new apps (3), unaware of updated privacy settings (2)}
  \item \textbf{Granular relationship-based sharing challenges (7)} 
  
    \textit{shared account (3), segment data recipient by human (4)}
  \item \textbf{Digital footprint management overload (15)}  
  
    \textit{cognitive challenge to remember (4), failure to track (7), too many apps (4)}
  \item \textbf{Evolving privacy preferences without historical management tools (20)}  
  
    \textit{change of preference (7), failure to track (7), manual delete (6)}
  \item \textbf{Interpersonal privacy boundary complications (8)}  
  
    \textit{private message disclosed to others (2), shared account (3), multiple people’s data (3)}
  \item \textbf{Identification protection challenges (19)}  
  
    \textit{false estimation of vulnerability (3), profile info management (12), overall digital identity (4)}
  \item \textbf{Inconsistent privacy customization capabilities (23)}  
  
    \textit{linked account (6), adjust privacy settings (6), default setting (2), unaware of updated privacy settings (2), not an available setting in this app (7)}
  \item \textbf{Accumulated data management deficiencies (17)}  
  
    \textit{all history data (7), evolving technology (1), manual delete (6), manual check-in (3)}
\end{itemize}

\subsection{Survey study}

Each code is followed by its description.

\begin{itemize}
  \item \textbf{Current strategy – passive} \textit{Give up to manage the nuance in content sharing}
  \item \textbf{Current strategy – all or nothing} \textit{(Resist to find the solution effective because) already not sharing anything}
  \item \textbf{Cross platform management} \textit{Needs and challenges to manage data cross applications, desire for a solution to satisfy all privacy management across multiple applications}
  \item \textbf{Cross device management} \textit{Management of data across multiple devices}
  \item \textbf{Multi-modality} \textit{Information sharing concern or control including non-text form (pictures, videos, location-sharing)}
  \item \textbf{Shared ownership} \textit{When share the ownership of the information with others, e.g., shared doc}
  \item \textbf{Other people’s privacy} \textit{Involves other people’s privacy; could be friends, families, co-workers, or strangers}
  \item \textbf{Data recipient – contextual timing} \textit{Want the data recipient to have access at a certain time and context}
  \item \textbf{Data recipient – group audience consideration} \textit{Discussion on consideration of different segments}
  \item \textbf{Data recipient – public vs private} \textit{Discussion about different mental models regarding public-accessible information versus private input}
  \item \textbf{High stakes data scenarios} \textit{Using privacy tools specifically when handling sensitive information (banking, healthcare, financial trading, therapy apps)}
  \item \textbf{Third party risk concern} \textit{Worry about privacy being compromised through the third party handling data}
  \item \textbf{Enterprise policy compliance} \textit{Privacy tool could help with organizational requirements}
  \item \textbf{Privacy feature discoverability} \textit{Struggles to find available privacy options within applications}
  \item \textbf{Expectation on data trade service frequency} \textit{Show less eagerness to contribute data if they only want to use the service once}
  \item \textbf{(AI) automation} \textit{Expected the technology to automate the process, save time, or make it effortless}
  \item \textbf{Unbiased AI} \textit{Emphasize the importance of having an unbiased technology}
  \item \textbf{(AI) accuracy – high confidence} \textit{Expected high accuracy from the technology; could also relate to the intelligence of AI}
  \item \textbf{(AI) accuracy – doubts} \textit{Doubtful attitude on whether the proposed technology can really handle information correctly}
  \item \textbf{Human–AI collaboration preference} \textit{Discussion around human factors vs. AI or human supervision on AI performance}
  \item \textbf{Small manual effort} \textit{Expected less effortful or manual work}
  \item \textbf{Centralized and comprehensive} \textit{Empathized that the technology should have a unified and comprehensive solution}
  \item \textbf{Break the wall of mystery data process} \textit{Described the technology as being able to make unknown known so as to empower the user}
  \item \textbf{Granular control preference} \textit{Wanting detailed, customizable privacy settings rather than one-size-fits-all solutions}
  \item \textbf{Hesitation to all or nothing strategy} \textit{Hesitate to delete all information}
  \item \textbf{Peace of mind} \textit{The AI-powered tool helps reduce their anxiety of managing privacy}
  \item \textbf{User agency – too intrusive} \textit{Resist considering the design because it is too intrusive: it might collect a lot of data or exert too much agency}
  \item \textbf{User agency – more tool autonomy} \textit{User expected the technology to provide more guidance or automation}
  \item \textbf{Post-sharing management} \textit{Describe use cases specific to the nature of post-sharing}
  \item \textbf{Pre-sharing risk assessment} \textit{Describe use cases specific to the nature of pre-sharing}
  \item \textbf{Dynamic preference adaptation – user} \textit{Expectation on updates in privacy control in response to changing life circumstances or mental models}
  \item \textbf{Dynamic preference adaptation – policy} \textit{Expect the technology to be compliant with up-to-date policy}
\end{itemize}


\end{document}